\pdfoutput=1
\documentclass[11pt,twoside,a4paper,cmspaper,final,collab]{cms-tdr}
\def\svnVersion{d59b1e0}\def\svnDate{2026/10/12}\def\cmsCernNoTag{CERN-EP-2026-057}\def\cmsCernDate{\today}\def\cmsMessage{Published in Physics Letters B as \href{http://dx.doi.org/10.1016/j.physletb.2026.140964}{\doi{10.1016/j.physletb.2026.140964}.}}
\begin{document}\cmsNoteHeader{EXO-24-011}

\newcommand{\CHI}{\ensuremath{\chi_{\text{dijet}}}\xspace}
\newcommand{\YBOOST}{\ensuremath{y_\text{boost}}\xspace}
\newcommand{\intlumi} {\ensuremath{138\fbinv}\xspace}
\newcommand{\mjj} {\ensuremath{M_{\mathrm{jj}}}\xspace}
\newcommand{\ptave}{\ensuremath{\langle \pt \rangle}\xspace}
\newcommand{\QBHMC} {{\textsc{qbh}}\xspace}
\ifthenelse{\boolean{cms@external}}{\providecommand{\cmsLeft}{upper\xspace}}{\providecommand{\cmsLeft}{left\xspace}}
\ifthenelse{\boolean{cms@external}}{\providecommand{\cmsRight}{lower\xspace}}{\providecommand{\cmsRight}{right\xspace}}

\cmsNoteHeader{EXO-24-011}
\title{Measurement of dijet angular distributions and search for beyond the standard model physics in proton-proton collisions at \texorpdfstring{$\sqrt{s} = 13\TeV$}{sqrt(s) = 13 TeV}}

\date{\today}

\abstract{
A measurement is presented of dijet angular distributions in proton-proton collisions at $\sqrt{s} = 13\TeV$, using data collected with the CMS detector at the CERN LHC and corresponding to an integrated luminosity of 138\fbinv. For the first time, the dijet angular distributions, corrected for detector effects, are compared with the predictions of perturbative quantum chromodynamics at next-to-next-to-leading order, including next-to-leading-order electroweak corrections. While data are generally found to be in agreement with predictions, a small difference in shape of the normalized distributions is seen for dijet masses ranging from 2.4 to 4.8\TeV and above 6\TeV. The distributions are used to search for proposed signatures of quark compositeness, extra spatial dimensions, quantum black holes, dark-matter mediators, axion-like particles, and anomalous gluon couplings. The most stringent limits to date are set for most of these scenarios. Quark contact interactions are excluded at 95\% confidence level (\CL) up to a scale of 17 (37)\TeV for destructive (constructive) interference in a benchmark scenario, valid to next-to-leading order in quantum chromodynamics, and in which only left-handed quarks participate. The coupling of axion-like particles to the gluon, $c_{\text{g}}/f_{\text{a}}$, is constrained to be lower than 0.42$\TeV^{-1}$ at 95\% \CL. The anomalous triple-gluon coupling, $C_{\text{G}}/\Lambda^2$, in a standard model effective field theory is constrained to be lower than 0.0076$\TeV^{-2}$ at 95\% \CL. }

\hypersetup{%
pdfauthor={CMS Collaboration},%
pdftitle={Measurement of dijet angular distributions and search for beyond the standard model physics in proton-proton collisions at 13 TeV},
pdfsubject={CMS},%
pdfkeywords={CMS, QCD, contact interactions, extra dimensions, black holes, dark matter, axions, effective field theory}}

\maketitle 

\section{Introduction}

Pairs of highly energetic jets of particles (dijets) are produced through pointlike scattering of quarks and gluons (partons) in proton-proton (pp)
collisions.  Measurements of dijet production at the CERN LHC allow the testing of parton interactions predicted by quantum chromodynamics (QCD)
at the highest energy scales with unprecedented precision~\cite{CMS:2012ftr,ATLAS:2013jmu,CMS:2017jfq,ATLAS:2017ble,CMS:2023fix}. As an attempt to
address open questions not explained by the standard model (SM) of particle physics, many theories beyond the SM (BSM) have been proposed. Several of
these models predict new interactions of quarks and gluons, making dijet production sensitive to signatures of BSM physics, including quark
compositeness~\cite{Terazawa:1980,CI-1,CI-2}, extra dimensions~\cite{ADD1,ADD2}, quantum black holes (QBHs)~\cite{BH1,Banks:1999gd,Emparan:2000rs,BH2,BH3},
dark-matter (DM) mediators~\cite{Abdallah:2015ter, Boveia:2016mrp}, axion-like particles (ALPs)~\cite{Gavela:2019cmq}, and anomalous gluon couplings in an effective field theory (EFT)~\cite{Buchmuller:1985jz}.

\begin{figure*}[htbp]
  \centering
  \includegraphics[width=0.24\textwidth]{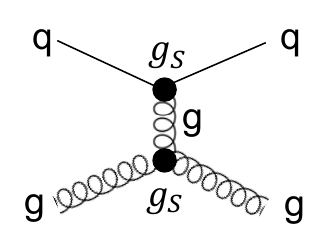}
  \includegraphics[width=0.24\textwidth]{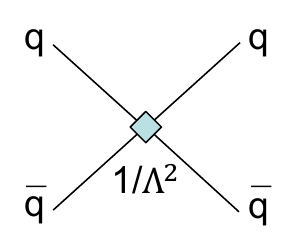}
  \includegraphics[width=0.24\textwidth]{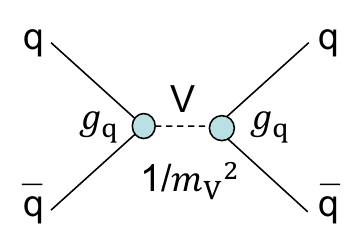}
  \includegraphics[width=0.24\textwidth]{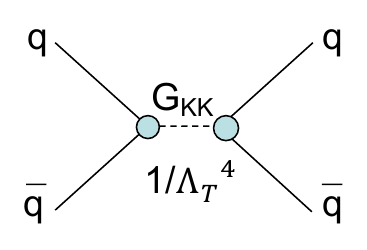} \\
  \includegraphics[width=0.24\textwidth]{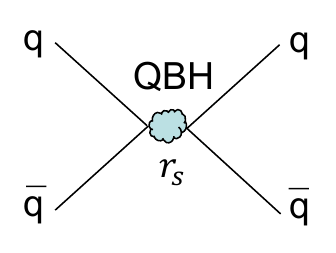}
  \includegraphics[width=0.24\textwidth]{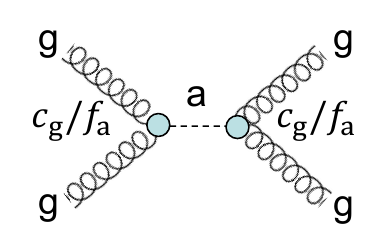}
  \includegraphics[width=0.24\textwidth]{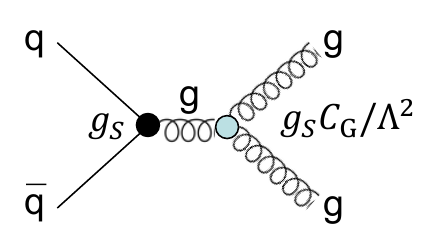}
  \caption{Example diagrams of dijet production processes in QCD and BSM physics. 
    Upper row: $t$-channel quark-gluon scattering in QCD (left), quark-quark scattering via a BSM contact interaction (center left), 
    quark pair production through an $s$-channel BSM resonance (center right), and quark pair production through gravitons in models with extra spatial dimensions (right). 
    Lower row: quark pair production through a quantum black hole (left), gluon pair production through an axion-like particle (center), 
    and gluon pair production with a dimension-6 effective field theory anomalous gluon coupling operator (right).
    A description of the symbols in the diagrams is given in Section~\ref{sec:bsm}.}
  \label{fig:feynman}
\end{figure*}

Example processes of dijet production through QCD and BSM interactions are depicted in Fig.~\ref{fig:feynman}.  At leading order (LO), dijet
production in $\Pp\Pp$ collisions has significant contributions from $t$-channel scattering (an example diagram with strong coupling $g_S$ is shown in Fig.~\ref{fig:feynman}, upper row, left). The BSM processes, however, can be
dominated by $s$-channel production, as well as four-fermion contact interactions (CIs).  Depending on the masses and couplings of BSM particles, they can manifest
themselves as resonances in pair production, or nonresonant modification of the dijet cross section as a function of the invariant mass and
scattering angle.  Resonant production of BSM particles has been strongly constrained by searches in the dijet invariant mass spectrum by the ATLAS and
CMS Collaborations~\cite{ATLAS:2019fgd,CMS:2019gwf}, and nonresonant enhancements of dijet production have been probed in measurements sensitive to the angular
distribution of jets~\cite{CMS:2021yzl,CMS:2018ucw,ATLAS:2017eqx}.

In searches for nonresonant effects from BSM phenomena in dijet production, it is crucial to have a precise description of the SM expectation.
Exploiting the huge phase space available for dijet cross sections, constraints can be imposed on the strong coupling constant and the
parton distribution functions (PDFs), as performed with triple-differential measurements in Refs.~\cite{CMS:2017jfq,CMS:2023fix}.
Those measurements are reported as functions of the dijet mass (\mjj), the Lorentz boost of the dijet center-of-mass (CM) system with respect to the lab system $\YBOOST = \abs{y_1+y_2}/2$, and $\abs{y_1-y_2}$, where $y_1$ and $y_2$ are the rapidities of the two jets with highest transverse momenta \pt (referred to as the two leading jets in the following).
A combined fit, also
including parameters representing contributions from CIs, has been performed using double-differential inclusive jet
measurements as a function of jet \pt and absolute rapidity $\abs{y}$ in Ref.~\cite{CMS:2021yzl}.

For the analysis presented in this Letter, to maximize the sensitivity to BSM effects, we focus on a much reduced phase space at large \mjj and low values of $\YBOOST$.
It is advantageous to choose an observable that minimizes the sensitivity to PDFs and the strong
coupling constant, which motivates the use of the normalized dijet angular distribution that is defined as
$\CHI= \exp(\abs{y_1-y_2})$. 
For collinear massless parton scattering, \CHI takes the form $\CHI = (1+ \cos{\theta^*})/(1 - \cos{\theta^*})$, where $\theta^*$ is the polar scattering angle in the parton-parton CM frame.
In contrast, Rutherford scattering, where only $t$-channel scattering contributes to the partonic cross section, is independent of \CHI.
The QCD dijet production cross section is almost independent of \CHI, apart from a small deviation at low
values of \CHI originating from subdominant contributions of $s$-channel QCD processes and higher-order corrections.  In contrast, many signatures of BSM physics would manifest
themselves as an excess of events at small values of \CHI.  Hence, the normalized distribution provides sensitivity to the scattering dynamics with minimal PDF dependence,
as the angular distributions of the dominant underlying QCD processes of
$\PQq\Pg\to \PQq\Pg$, $\PQq\PAQq$(\PQq')${}\to\PQq\PAQq$(\PQq'), and $\Pg\Pg\to\Pg\Pg$ are similar~\cite{Combridge:1983jn}.
The \CHI distribution is measured in bins of \mjj to enhance the sensitivity to many signatures of BSM physics whose contribution to the dijet cross section rises with \mjj.

In this Letter, our previous measurement of dijet angular distributions~\cite{CMS:2018ucw} is extended to higher dijet masses, using a data sample collected in 2016--2018 by the CMS experiment at $\sqrt{s} = 13\TeV$, 
and corresponding to an integrated luminosity of \intlumi~\cite{CMS-PAS-LUM-17-003, CMS-PAS-LUM-17-004, CMS-PAS-LUM-18-002}. The strategy of the analysis is two-fold.  First, we search for evidence
of BSM physics in the dijet angular distributions as measured by the detector (detector level), considering models with CIs, extra dimensions, QBHs, and DM mediators. New constraints are also placed on models with
ALPs~\cite{Gavela:2019cmq} and on anomalous gluon coupling in an EFT~\cite{Goldouzian:2020wdq}.  Second, we present a new
measurement of the dijet angular distributions, unfolded for detector effects  (particle level).  For the first time, the dijet angular distributions corrected for detector effects are compared to a
next-to-next-to-LO (NNLO) perturbative QCD prediction including next-to-LO (NLO) electroweak (EW) corrections~\cite{Currie:2017eqf}. 
Tabulated results can be found in the HEPData record for this analysis~\cite{hepdata}.

\section{The CMS detector}

The CMS apparatus~\cite{CMS:2008xjf,CMS:2023gfb} is a multipurpose, nearly hermetic detector, designed to trigger on~\cite{CMS:2020cmk,CMS:2016ngn,CMS:2024aqx} and identify electrons, muons, photons, and (charged and neutral) hadrons~\cite{CMS:2020uim,CMS:2018rym,CMS:2014pgm}. Its central feature is a superconducting solenoid of 6\unit{m} internal diameter, providing a magnetic field of 3.8\unit{T}. Within the solenoid volume are a silicon pixel and strip tracker, a lead tungstate crystal electromagnetic calorimeter (ECAL), and a brass and scintillator hadron calorimeter, each composed of a barrel and two endcap sections. Forward calorimeters extend the pseudorapidity coverage provided by the barrel and endcap detectors. Muons are reconstructed using gas-ionization detectors interleaved with the layers of the steel flux-return yoke outside the solenoid. More detailed descriptions of the CMS detector, together with a definition of the coordinate system used and the relevant kinematic variables, can be found in Refs.~\cite{CMS:2008xjf,CMS:2023gfb}.
Events of interest are selected using a two-tiered trigger system. The first level (L1), composed of custom hardware processors, uses information from the calorimeters and muon detectors to select events at a rate of around 100\unit{kHz} within a fixed latency of 4\mus~\cite{CMS:2020cmk}. The second level, known as the high-level trigger (HLT), consists of a farm of processors running a version of the full event reconstruction software optimized for fast processing, and reduces the event rate to a few kHz before data storage~\cite{CMS:2016ngn,CMS:2024aqx}.

\section{Event reconstruction and selection}

Events are reconstructed using the particle-flow
algorithm~\cite{Sirunyan:2017ulk} to identify and reconstruct
individual particles from each collision by combining information from
all CMS subdetectors.  Identified particles include charged
hadrons, neutral hadrons, electrons, muons, and photons. The particles are
clustered into jets using the anti-\kt algorithm~\cite{antikt:2008}
with a distance parameter of $R=0.4$,
as implemented in the \FASTJET package~\cite{Cacciari:2011ma}.
The jet energies are corrected for the
nonlinear and nonuniform response of the calorimeters through corrections
obtained from data and Monte Carlo (MC) simulations~\cite{Khachatryan:2016kdb}.
The distance parameter of $R=0.4$ is chosen, as the jet calibration is the dominant source of systematic uncertainty in this measurement and is most precisely known for this value of $R$.

Events are selected online using a variety of different jet triggers based on the highest jet \pt or the scalar sum of the \pt of all jets in the event (\HT).
The selection threshold varied over the course of the
data-taking period, ranging in the HLT between 450 and 550\GeV, and 900 and 1050\GeV for \pt and \HT, respectively.

In the offline event selection, events with at least two reconstructed jets are selected.
Spurious jets from noise are rejected by applying
loose quality criteria~\cite{CMS-PAS-JME-16-003} to jet properties.
Anomalous events due to a variety of reconstruction failures, detector malfunctions, or noncollision backgrounds are rejected by event filters that are designed to identify more than 85--90\% of the spurious events, with a mistagging rate less than 0.1\%~\cite{CMS:2019ctu}.
The primary vertex~\cite{CMS:2014pgm} is taken to be the reconstructed vertex corresponding to the hardest scattering in the event, evaluated using tracking information alone, as described in Section 9.4.1 of Ref.~\cite{CMS-TDR-15-02}.
For each event, a reconstructed primary vertex is required that lies within
$\pm$24\unit{cm} of the detector center along the beam line, and within
2\unit{cm} of the detector center in the plane transverse to the beam.

The two leading jets
are used to measure the \CHI distributions
in several regions of \mjj.
The \mjj regions considered are listed in Table~\ref{tab:events}.
The width of the \mjj regions is larger than the experimental resolution,
and the highest \mjj region was defined to maximize the
expected sensitivity to the BSM signals considered, while
guaranteeing at least 50 expected events in that region.
The phase space for this analysis
is defined by events with $1<\chi_\text{dijet}<16$ and
$\abs{ \YBOOST }<1.11$.
This selection restricts the rapidities $\abs{y_1}$ and $\abs{y_2}$ of
the two highest \pt jets to be less than 2.5, \ie, within the acceptance of the tracking detector.
The dijets with the highest (leading) and second-highest (sub-leading) jet \pt are required to have \pt values above 500 and 200\GeV, respectively.
The trigger efficiency exceeds 99\% over the entire analysis phase space.
The observed number of events, $N_{\text{dijet}}$, after selection in each \mjj bin is listed in Table~\ref{tab:events}.
The highest value of \mjj observed in the data is 8.3\TeV.

\begin{table}[b]
\centering
\topcaption{\label{tab:events} Observed number of events, $N_{\text{dijet}}$, after selection in each $M_{\text{jj}}$ bin.}
\renewcommand{\arraystretch}{1.3}
\begin{tabular}{cc} 
$M_{\text{jj}}$ bin [{\TeVns}] & $N_{\text{dijet}}$ \\
\noalign{\vskip 0.5mm} 
\hline
\begin{tabular}{r} 2.4--3.0 \\ 3.0--3.6 \\ 3.6--4.2 \\ 4.2--4.8 \\ 4.8--5.4 \\ 5.4--6.0 \\ 6.0--7.0 \\ $>$7.0 \end{tabular} & 
\begin{tabular}{r} 1\,363\,868 \\ 283\,202 \\ 68\,232 \\ 17\,731 \\ 5\,040 \\ 1\,423 \\ 433 \\ 50 \end{tabular} \\
\end{tabular}
\end{table}

The \CHI distributions obtained after event reconstruction and selection,
called detector level in the following, are normalized to unity in each \mjj region.
These distributions,
\begin{equation}
\frac{1}{N_{\text{dijet}}}\ \frac{\text{d}N_{\text{dijet}}}{\text{d}\CHI},
\end{equation}
are then analyzed in two ways. First, signatures of BSM physics are searched for at the 
detector level. Second, the distributions are unfolded for detector effects to obtain the dijet 
angular distributions at what is referred to as the particle level,
\begin{equation}
\frac{1}{\sigma_{\text{dijet}}}\ \frac{\text{d}\sigma_{\text{dijet}}}{\text{d}\CHI},
\end{equation}
where $\sigma_\text{dijet}$ is the corrected total dijet cross section in the \mjj region.

Fluctuations in jet response arising from the jet \pt resolution of the
detector can cause lower-energy jets to be misidentified as one of the two leading jets.
Such fluctuations can also produce bin-to-bin migrations in the \CHI and \mjj regions.
The response matrix, obtained by smearing particle-level jets with a detector response parameterization (discussed in Section~\ref{sec:theory}), encodes the probabilities that an event in a given bin $j$ at the particle level is reconstructed in a bin $i$ at the detector level.
The response matrix enables both forward folding of theoretical predictions to the detector level for BSM searches and backward unfolding of the raw data distributions to the particle level for the measurement.
The search for BSM physics utilizes the detector-level distributions to avoid uncertainties related to the inversion of the response matrix and also because the measured data counts can be treated as uncorrelated among all bins in the statistical analysis, unlike after unfolding to particle level.

\section{The QCD predictions and uncertainties}
\label{sec:theory}

We compute predictions of the normalized dijet angular distribution from perturbative QCD at NNLO using \textsc{NNLOJET}~\cite{Currie:2017eqf} in the
\textsc{fastnlo} framework~\cite{fastnlo}. The same jet clustering algorithm used at the detector level is applied to the outgoing partons in the NNLO
prediction.  Following recommendations outlined in Ref.~\cite{Currie:2017eqf}, the dijet invariant mass of the two leading jets, \mjj, is
chosen as the central reference value for both the factorization scale, $\mu_{\mathrm{F}}$, and renormalization scale, $\mu_{\mathrm{R}}$, in the perturbative
calculation: $\mu_{\mathrm{F}} = \mu_{\mathrm{R}} = \mjj$. For comparison to our previous publication~\cite{CMS:2018ucw} using predictions at NLO, the average \pt
of the two leading jets, \ptave, is used as an alternative: $\mu_{\mathrm{F}}$ = $\mu_{\mathrm{R}}$ = \ptave. Since $\mjj = 2\ptave\cosh(\abs{y_1 -y_2}/2)$, the two
scale choices, \mjj and \ptave, coincide up to a factor of two for $\CHI = \exp(\abs{y_1 -y_2})$ approaching unity, but differ for larger values of \CHI.

The two scale choices exhibit different perturbative convergence behavior, which is usually associated with
predictions at some order $i$ that successively include the next higher-order prediction within the estimated uncertainty. In particular, $K$ factors,
$K_i = \sigma_{\text{N}^{i+1}\text{LO}}/\sigma_{\text{N}^{i}\text{LO}}$, are expected to lie around unity within scale variations and to be closer to unity with increasing $i$. 
In addition, scale variation uncertainty bands are expected to overlap and decrease in size with increasing $i$.  This is fulfilled for
our central scale choice of \mjj, but not for \ptave, where the scale dependence at NNLO can be as large as that at NLO.  
Due to a better perturbative convergence behavior, we choose \mjj as the central scale, while \ptave is
employed only as an alternative to derive an additional estimate for the scale dependence of the normalized dijet angular distribution.

For all simulated samples, the PDFs are taken from the NNPDF3.1 set~\cite{NNPDF:2017mvq} at NNLO.
As an alternative PDF, and for comparison to our previous publication~\cite{CMS:2018ucw}, the CT14 sets~\cite{Dulat:2015mca} at NNLO and NLO were also studied.

Electroweak corrections for dijet production~\cite{Dittmaier:2012kx} have
a nonnegligible effect on the normalized $\chi_\text{dijet}$
distributions, requiring corrections of 1\% (5\%) at low
(high) \mjj.

We evaluated the impact of nonperturbative effects due to the
hadronization and multiple parton interactions on the QCD predictions
using \PYTHIA 8.212 with the CUETP8M1
tune~\cite{Skands:2014pea,Khachatryan:2015pea} and {\HERWIG{}++}
2.7.1~\cite{herwig:2008} with tune EE5C~\cite{Seymour:2013qka}.
Particle-level jets are computed from the four-vectors
of generated stable particles
(lifetime $c\tau > 1\cm$, including muons, excluding neutrinos).
The effects are found to be
negligible in both MC generators.
Therefore, we are able to directly compare the
data corrected to the particle level with the parton-level theoretical
predictions.

The uncertainty related to the choices of the
$\mu_{\mathrm{F}}$ and $\mu_{\mathrm{R}}$ scales are evaluated
following the proposal in Refs.~\cite{Cacciari:2003fi,Banfi:2010xy} by varying the
default choice of scales in the following 6 combinations:
$(\mu_{\mathrm{F}}/\mjj$, $\mu_{\mathrm{R}}/\mjj)$ = $(1/2, 1/2)$, $(1/2, 1)$,
$(1,1/2)$, $(1, 2)$, $(2, 1)$, and $(2, 2)$. 
The uncertainty in PDFs is determined using the 100
MC replicas provided in NNPDF3.1~\cite{Ball:2014uwa}.
The uncertainty in the strong
coupling constant has a negligible impact on the normalized
\CHI distribution.
The NNLO calculation has a statistical uncertainty from the MC integration.
A summary of the leading
systematic uncertainties in the theoretical predictions is shown in
Table~\ref{tab:systh}.
The two central scale choices display a considerably different perturbative convergence and residual scale uncertainty.
For the BSM search, we account for the difference between the scale choices \mjj and \ptave as an additional uncertainty, since they are not covered by the scale uncertainty from variations of the scale around the central choice $\mu_{\mathrm{F}} = \mu_{\mathrm{R}} = \mjj$.

\begin{table}[htb]
\centering
\topcaption{\label{tab:systh} Summary of the leading theoretical uncertainties in the normalized \CHI distributions.
The full analysis takes into account correlations between \CHI bins; however, this table shows the contributions in the lowest \CHI bin for two representative $M_{\text{jj}}$ bins. This illustrates the magnitude of the contributions for these $M_{\text{jj}}$ bins.}
\renewcommand{\arraystretch}{1.3}
\begin{tabular}{lrr}
$M_{\text{jj}}$ bin [{\TeVns}]              & 3.0--3.6 & $>$7.0 \\
\noalign{\vskip 0.5mm} 
\hline
$\mu_{\mathrm{R,F}}$ variations (6 points)   & $_{-0.4\%}^{+0.2\%}$ & $_{-1.5\%}^{+2.7\%}$ \\
$\mu_{\mathrm{R,F}}$ choice ($M_{\text{jj}}$ vs. $\ptave$) & 2.5\% & 9.6\% \\
PDF (NNPDF3.1 NNLO)                    & 0.1\%  & 5.5\% \\
Theory NNLO statistical                & 0.5\%  & 4.4\% \\
Nonperturbative effects               & $<$1\% & $<$1\% \\\hline
Total theoretical                      & 3.5\%  & 12.2\% \\
\end{tabular}
\end{table}

To map the above particle-level predictions to the detector level, we make use of MC event generators and a detailed simulation of the CMS detector using \GEANTfour~\cite{Agostinelli:2002hh}.
Samples of QCD multijet production are generated at LO with \PYTHIA 8.212~\cite{pythia8:2015}  with the CP5 tune~\cite{Sirunyan:2019dfx}.
An alternative sample of QCD multijet production is generated at LO with \MGvATNLO v2.6.5~\cite{Alwall:2014hca} with up to four additional jets in the matrix element computation.
The sample is binned in the scalar sum of all jet transverse momenta \HT.
The \MGvATNLO generator is interfaced with \PYTHIA  for the parton showering and hadronization.
To avoid double-counting of jets from the matrix element calculations and the parton shower, jets are matched using the MLM scheme~\cite{Alwall:2007fs}.
A second alternative sample of QCD multijet production is obtained using \HERWIG 7.1.4~\cite{Bellm:2015jjp,herwig:2008}
at LO.

To predict the background in the search for BSM physics, a detector-level prediction is computed starting from the prediction at NNLO in QCD with EW corrections.
A response matrix is applied, mapping the particle-level \mjj and \CHI distributions onto the measured values.
The response matrix is obtained from the \PYTHIA MC event generator.
Detector-level jets are determined from particle-level jets by smearing their \pt, using a double-sided Crystal Ball parameterization~\cite{crystalball,Gaiser:1982yw} of the response that varies as a function of \pt and $\eta$.
The parameterization takes into account the full jet energy resolution, including non-Gaussian tails estimated using detector simulation, and verified in data~\cite{Khachatryan:2016kdb}.

\section{The BSM predictions and uncertainties}
\label{sec:bsm}

A common signature of quark compositeness~\cite{Terazawa:1980,CI-1,CI-2}
is the appearance of new interactions between quark constituents
at some characteristic mass scale, $\Lambda$, that is much larger than the
quark masses.  At energies well below $\Lambda$, these
interactions are approximated with a CI
characterized by a four-fermion coupling (an example diagram is shown in Fig.~\ref{fig:feynman}, upper row), described by the effective
Lagrangian~\cite{CI-1,CI-2}:
\ifthenelse{\boolean{cms@external}}
{
\begin{multline*}
\mathcal{L}_{qq}=\frac{2\pi}{\Lambda^2} \Big[
    \eta_{\mathrm{LL}}({\overline{q}}_{\mathrm{L}}\gamma^{\mu}q_{\mathrm{L}})(\overline{q}_{\mathrm{L}}\gamma_{\mu}q_{\mathrm{L}})
    + \\ 
    \eta_{\mathrm{RR}}(\overline{q}_{\mathrm{R}}\gamma^{\mu}q_{\mathrm{R}})(\overline{q}_{\mathrm{R}}\gamma_{\mu}q_{\mathrm{R}})
    +2\eta_{\mathrm{RL}}(\overline{q}_{\mathrm{R}}\gamma^{\mu}q_{\mathrm{R}})(\overline{q}_{\mathrm{L}}\gamma_{\mu}q_{\mathrm{L}})
  \Big],
\end{multline*}
}
{
\begin{equation*}
  \mathcal{L}_{qq}=\frac{2\pi}{\Lambda^2} \left[
    \eta_{\mathrm{LL}}({\overline{q}}_{\mathrm{L}}\gamma^{\mu}q_{\mathrm{L}})(\overline{q}_{\mathrm{L}}\gamma_{\mu}q_{\mathrm{L}})
    +\eta_{\mathrm{RR}}(\overline{q}_{\mathrm{R}}\gamma^{\mu}q_{\mathrm{R}})(\overline{q}_{\mathrm{R}}\gamma_{\mu}q_{\mathrm{R}})
    +2\eta_{\mathrm{RL}}(\overline{q}_{\mathrm{R}}\gamma^{\mu}q_{\mathrm{R}})(\overline{q}_{\mathrm{L}}\gamma_{\mu}q_{\mathrm{L}})
  \right],
\end{equation*}
}
where the subscripts $\mathrm{L}$ and $\mathrm{R}$ refer to the left and right chiral projections of
the quark fields, respectively, and $\eta_{\mathrm{LL}}$, $\eta_{\mathrm{RR}}$, and $\eta_{\mathrm{RL}}$ are
integers with chosen values of 0, ${+}1$, or ${-}1$. The various combinations of
$\left(\eta_{\mathrm{LL}},\,\eta_{\mathrm{RR}},\,\eta_{\mathrm{RL}}\right)$ correspond to
different CI models. The investigated CI scenarios with color-singlet
couplings between quarks use the $(\eta_{\mathrm{LL}},\,\eta_{\mathrm{RR}},\,\eta_{\mathrm{RL}})$ values shown in Table~\ref{tab:lambdas}.

\begin{table}[htb]
\centering
\topcaption{\label{tab:lambdas} The different combinations of $(\eta_{\text{LL}},\,\eta_{\text{RR}},\,\eta_{\text{RL}})$ used
      in the search for CIs between quarks and the corresponding mass scale name.}
      \renewcommand{\arraystretch}{1.3}
  \begin{tabular}{lc} 
    $\Lambda$ & $(\eta_{\text{LL}},\,\eta_{\text{RR}},\,\eta_{\text{RL}})$ \\
    \hline
    $\Lambda^{\pm}_{\text{LL}}$    & $(\pm 1,\hphantom{\pm}0,\hphantom{\pm}0)$ \\
    $\Lambda^{\pm}_{\text{RR}}$    & $(\hphantom{\pm}0,\pm 1,\hphantom{\pm}0)$ \\
    $\Lambda^{\pm}_{\text{VV}}$    & $(\pm 1,\pm 1,\pm 1)$ \\
    $\Lambda^{\pm}_{\text{AA}}$    & $(\pm 1,\pm 1,\mp 1)$ \\
    $\Lambda^{\pm}_{\text{(V--A)}}$ & $(\hphantom{\pm}0,\hphantom{\pm}0,\pm 1)$ \\
  \end{tabular}
\end{table}

We note that the models with positive (negative) $\eta_{\text{LL}}$ or $\eta_{\text{RR}}$ lead to
destructive (constructive) interference with the QCD terms and a lower (higher) dijet cross section.
In all CI models discussed in this paper, NLO QCD corrections are
employed to calculate the cross sections. In pp collisions, the $\Lambda^{\pm}_{\text{LL}}$ and
$\Lambda^{\pm}_{\text{RR}}$ models result in identical tree-level cross sections and NLO corrections,
and consequently lead to the same sensitivity. For
$\Lambda^{\pm}_{\text{VV}}$, $\Lambda^{\pm}_{\text{AA}}$, and
$\Lambda^{\pm}_{(\text{V-A})}$ the CI predictions are
identical at tree level, but exhibit different NLO corrections and yield different sensitivity.
To calculate the CI terms and the interference between the
CI and QCD terms at NLO in QCD, the \textsc{cijet}~1.0 program~\cite{contactnlo} is employed.
Scale and PDF uncertainties in the CI terms and QCD processes are treated as correlated.

Another BSM physics scenario considered is a DM mediator particle that couples to DM particles and quarks.
Depending on the couplings of such a mediator particle to DM and quarks, it can be observed as a narrow or wide resonance or even a nonresonant dijet signature.
In a simplified model of interactions between DM and quarks~\cite{Abdallah:2015ter, Boveia:2016mrp}, a DM mediator with unknown mass, $m_{\text{med}}$, is assumed to be a spin-1 particle (vector $\mathrm{V}$ or axial-vector $\mathrm{A}$) and to decay only to pairs of quarks or DM particles (an example diagram for a vector mediator with mass $m_{\text{V}}$ is shown in Fig.~\ref{fig:feynman}, upper row).
The DM mediator has a universal quark coupling, $g_{\text{q}}$, and a DM coupling, $g_{\text{DM}}$.
The DM particles have mass, $m_{\text{DM}}$, which is set to 1\GeV in this analysis.
For models with $g_{\text{q}} \approx g_{\text{DM}} \approx 1$ and $2 m_{\text{DM}} < m_{\text{med}}$, the total width of the mediator is dominated by the width of the quark decay channel due to the large number of possible quark flavors and colors.
Therefore, the sensitivity to the parameters of interest for this search, $m_{\text{med}}$ and $g_{\text{q}}$, has little dependence on $m_{\text{DM}}$. 
Interactions of a DM vector and axial-vector mediator in this model are generated using \textsc{MadDM} version 2.0.6~\cite{Backovic:2013dpa,Backovic:2015cra}.
Interference with the SM production is not included in the simplified DM mediator model.
As the dijet mass range under study extends beyond $m_{\text{med}}=6.2\TeV$ up to which this search is sensitive, we do not expect the interference contributions to modify the phenomenology.
Scenarios with $g_{\text{q}}<0.5$ that predict resonances with a width lower than 10\% of the DM mediator mass have been excluded in dijet resonance searches~\cite{ATLAS:2019fgd,CMS:2019gwf} across a large range of $m_{\text{med}}$.
Therefore, scenarios with $g_{\text{q}}>0.5$ are of particular interest in this search.

The Arkani-Hamed--Dimopoulos--Dvali (ADD) model~\cite{ADD1,ADD2}
of compactified large extra dimensions (EDs) provides a possible
solution to the SM hierarchy problem.
In pp collisions at the LHC, it predicts signatures of virtual graviton exchange
that result in a nonresonant enhancement of dijet production (an example diagram is shown in Fig.~\ref{fig:feynman}, upper row), whose angular
distribution differs from the predictions of QCD.
For the ADD model, two parameterizations for virtual graviton exchange are considered:
Giudice--Rattazzi--Wells (GRW)~\cite{GRW} and Han--Lykken--Zhang
(HLZ)~\cite{HLZ}.
In the GRW convention, the sum over the Kaluza--Klein
graviton excitations, $G_{\text{KK}}$, in the effective field theory is regulated by a
single cutoff parameter, $\Lambda_{\text{T}}$.
In the HLZ convention, the effective
theory is described in terms of two parameters:
the string scale, $M_{\text{S}}$ (by convention used as the cutoff);
and the number of extra spatial dimensions, $n_{\text{ED}}$.
The parameters $M_{\text{S}}$ and $n_{\text{ED}}$ are directly related to
$\Lambda_{\text{T}}$~\cite{landsberg}.  We consider scenarios with 2--6 EDs.
The case of $n_{\text{ED}}=1$ is not considered, since it would require an
ED roughly the size of the solar system; the gravitational potential
at these distances would be noticeably modified, and this case is therefore excluded.
The case of $n_{\text{ED}}=2$ is special in
the sense that the relation between $M_{\text{S}}$ and $\Lambda_{\text{T}}$ also
depends on the parton-parton CM energy $\sqrt{\hat{s}}$.
The ADD model predictions are calculated with \PYTHIA.

In models with large EDs, the fundamental Planck scale is assumed to be closer
to the EW scale, thereby implying the possibility of black hole production
at the LHC~\cite{BH1,Banks:1999gd,Emparan:2000rs,BH2,BH3}.
Quantum black holes that are produced with a mass close to the Planck scale decay
predominantly into dijets and can be studied using dijet angular
distributions~\cite{QBH,QBH1,QBH2,Kanti:2002ge}.
The QBH production is studied within the framework of the ADD model, with $n_{\text{ED}}=6$ (ADD6),
and the Randall--Sundrum model with $n_{\text{ED}}=1$ (RS1)~\cite{RS1,RS2} (an example diagram is shown in Fig.~\ref{fig:feynman}, lower row).
In these models, the QBH production cross section
is typically described by the classical geometrical cross section, $\sigma_\text{QBH} \approx \pi r_{\text{s}}^{2}$,
where $r_{\text{s}}$ is the Schwarzschild radius of the QBH. The Schwarzschild radius depends on
the mass of the QBH, the Planck scale ($M_{\text{P}}$), and the number of spatial dimensions.
Since QBHs are produced with a mass threshold close to the Planck scale, we set the minimum QBH mass, $M_{\text{QBH}}$, equal to $M_{\text{P}}$ for simplicity.
The \QBHMC 3.0 generator~\cite{QBHGenRef} is used for the predictions.
In this implementation, the initial angular momentum of the QBH is assumed to be due entirely to the spin states of the incoming partons, and the possibility of an initial small orbital angular momentum due to a nonzero impact parameter of the colliding partons is ignored.

Axion-like particles~\cite{Georgi:1986df,Choi:1986zw} appear
in extensions of the SM that could potentially solve the strong $CP$ problem~\cite{Peccei:1977hh}
and/or explain the nature of DM.
Production of ALPs at the LHC can be described by an EFT with a linear realization of electroweak symmetry breaking,
adding an ALP with a mass $m_\text{a}$ to the SM with a new interaction at a characteristic high energy scale, $f_\text{a}$~\cite{Brivio:2017ije} (an example diagram is shown in Fig.~\ref{fig:feynman}, lower row).
In the limit $m_\text{a} \ll f_\text{a}$, ALP couplings to SM bosons are purely derivative in the sense that the coupling term in the Lagrangian only contains the derivative of the ALP field.
As a result, ALP interactions with SM bosons have cross sections that are enhanced at CM energies much larger than $m_\text{a}$.
The coupling to gluons is $g_{\mathrm{agg}} = 4 c_\text{g}/f_\text{a}$, where $c_\text{g}$ is the gluon operator coefficient.
Within the range of validity of the EFT, enhancement of $s$-channel gluon pair production can be probed in dijet events~\cite{Gavela:2019cmq}.
Predictions are computed using \MGvATNLO 2.4.3 with the \textsc{FeynRules} ALP\_linear\_UFO implementation of an EFT for a light ALP~\cite{Brivio:2017ije}. 
The calculation is performed at LO with up to one extra jet in the matrix element calculations, using the MLM matching~\cite{Alwall:2007fs}, and includes the interference with the SM production.
We study modifications of the ALP coupling to gluons $g_{\text{agg}}$, while all other ALP couplings are set to 0 and the ALP mass to 1\MeV, \ie, effectively massless compared to the CM energy.

More generally, dijet angular distributions can be used to constrain couplings in the SM EFT (SMEFT)~\cite{Goldouzian:2020wdq,Buchmuller:1985jz},
assuming that BSM physics lies at an energy scale much larger than the experimentally probed interaction energy.
The SMEFT Lagrangian includes higher-dimension operators involving SM
particles that are suppressed by powers of the BSM physics scale $\Lambda$.
A dimension-6 triple-gluon operator is introduced with a coefficient $C_\text{G}$ and is suppressed by $1/\Lambda^2$.
The operator modifies processes with gluons in the initial or final state (an example diagram is shown in Fig.~\ref{fig:feynman}, lower row).
Consistent with Ref.~\cite{Goldouzian:2020wdq}, predictions are computed using \MGvATNLO 2.4.3 with the \textsc{FeynRules} implementation TopEffTh of an EFT~\cite{Zhang:2010dr} at LO with up to two extra jets in the matrix element calculation, using the MLM matching, and no interference with SM production.
We study modifications of the triple-gluon coupling $C_\text{G}/\Lambda^2$, while all other anomalous couplings are set to 0.
The interference with the SM production of order $C_\text{G}/\Lambda^2$ was found to have no phenomenological relevance in dijet production up to NLO at the LHC energies~\cite{Hirschi:2018etq}.
The analysis reported here is mainly sensitive to contributions of order $C_\text{G}^2/\Lambda^4$.

To take into account the NNLO QCD and NLO EW corrections to the SM dijet production when probing the BSM
models, the cross section difference
$\sigma^{\text{QCD}}_{\text{NNLO+EW corr}} - \sigma^{\text{QCD}}_{\text{LO}}$
is evaluated for each \mjj and \CHI bin and added to the BSM
predictions that include LO SM QCD processes, namely ADD, QBH, DM, ALP, and SMEFT.
With this procedure, an SM+BSM prediction is
obtained where the QCD terms are corrected to NNLO including NLO EW corrections, while the BSM
terms are calculated at LO.
While the ADD and ALP BSM predictions (obtained by subtracting the predictions $\sigma^{\text{BSM+QCD}}_{\text{LO}} - \sigma^{\text{QCD}}_{\text{LO}}$) include the interference terms with the SM,
the QBH, DM, and SMEFT BSM predictions do not include such interference terms.
To avoid potential overestimation of sensitivity, we do not apply any $K$-factors, \ie, higher-order corrections, to the BSM contributions. We therefore do not assume that higher-order corrections for SM QCD interactions with qg, qq, and gg initial states apply to signals that modify only particular processes (those shown in Fig.~\ref{fig:feynman}).

Scale and PDF uncertainties in the CI predictions are taken into account.
In the ADD and QBH models, the scale and PDF uncertainties have a negligible
impact on the sensitivity, as the signals only appear in the highest mass bins, where the statistical
uncertainties dominate.
For the DM models, the effect on the acceptance due to the PDF uncertainty is found to be nonnegligible for $\mjj>m_{\text{med}}$ with DM mediators that have large mass and coupling.  For
example, for an axial-vector mediator with $m_{\text{med}} = 6\TeV$ and $g_{\text{q}} = 1.0$, which corresponds
to a resonance with $\Gamma/m_{\text{med}} = 50\%$, the uncertainty is 14\% in the $6.0<\mjj<7.0\TeV$ bin.

\section{Experimental uncertainties}

The normalized dijet angular distributions have reduced sensitivity to many experimental systematic effects compared to dijet mass \mjj or inclusive jet \pt cross section measurements.
Systematic uncertainties arise from the
jet energy scale (JES), the jet energy resolution (JER), triggering,
and the detector simulation and modeling used to construct the response matrix.
A summary of the size of leading experimental systematic uncertainties is provided in Table~\ref{tab:sys}.

\begin{table}[htb]
\centering
\topcaption{\label{tab:sys} Leading experimental uncertainties in the normalized \CHI distributions.
The uncertainties in the lowest \CHI bin for two $M_{\text{jj}}$ bins are shown as representative examples.
}
\renewcommand{\arraystretch}{1.3}
\begin{tabular}{lrr}
$M_{\text{jj}}$ bin [{\TeVns}] & 3.0--3.6 & $>$7.0 \\
\noalign{\vskip 0.5mm} 
\hline
Statistical                    & 0.7\%   & 25\%   \\
Jet energy scale               & 2.7\%   & 6.8\%  \\
Jet energy resolution (core)   & 0.1\%   & 2.0\%  \\
Jet energy resolution (tails)  & 0.5\%   & 4.7\%  \\
ECAL L1 trigger                & 0.3\%   & 2.8\%  \\
Response matrix modeling                       & 0.6\%   & 5.0\%  \\
Detector simulation            & 1.0\%   & 4.6\%  \\
Pileup                         & $<$0.1\% & $<$0.1\% \\ \hline

Total experimental             & 3.0\%   & 27\%   \\
\end{tabular}
\end{table}

The overall JES uncertainty is less than 1\%, and the variation of the JES as a function of
pseudorapidity in the phase
space of the analysis is less than 1\% per unit $\eta$~\cite{Khachatryan:2016kdb,jets2015}.  The JES uncertainties related to each step in the derivation of the \pt-
and $\eta$-dependent JES corrections are taken into account separately.
The correlations of the JES uncertainty sources among the \mjj ranges and \CHI bins are included.

The JER uncertainty in this measurement is evaluated by
changing the \pt-dependent width of the Gaussian core of the Crystal Ball parameterization
of the response by up to $\pm$5\%~\cite{Khachatryan:2016kdb,jets2015}, depending on
the jet pseudorapidity, and comparing the resultant
distributions before and after these changes.
The systematic uncertainty from the modeling of the tails of the jet resolution
is evaluated using a Gaussian function to parameterize the response,
and assigning as an uncertainty half of the difference
between the distributions determined from this Gaussian function and the nominal
function.

A source of uncertainty in the detector simulation used for the response matrix arises from the
use of a parameterized model to simulate the jet \pt resolution of the detector.
This uncertainty is estimated by comparing the \(\chi_\text{dijet}\) distributions obtained from this parametric model to those from a detailed simulation of the CMS detector using \GEANTfour~\cite{Agostinelli:2002hh}.
An additional systematic uncertainty is evaluated to account for the dependence of the response matrix on the physics model used to describe the dijet kinematics.
This uncertainty is determined by applying
the unfolding corrections (detailed in Section~\ref{sec:unfolding}) determined with \PYTHIA to smeared
$\chi_\text{dijet}$ distributions from \MGvATNLO
and \HERWIG
and comparing the results with the generated $\chi_\text{dijet}$ distributions.

During the 2016 and 2017 data-taking periods, a gradual shift in the timing of the inputs of the ECAL L1 trigger in the $\abs{\eta} > 2.0$ region caused a trigger inefficiency, denoted as ``prefiring''~\cite{CMS:2020cmk}.
Correction factors were computed from data and applied, with an effect of order $<$1\% in $2.0 < \abs{\eta} < 2.5$.
The uncertainty related to the correction for this effect is also taken into account.

The effect on the \CHI distributions from additional interactions in the same or adjacent
proton bunch crossings relative to the interaction of interest (pileup)
is studied by varying the pileup distribution in simulated samples, taking into account luminosity-related dependencies.

The statistical analysis of the data treats uncertainties separately; however, in tables and figures a quadratic sum of all contributions is presented.
While statistical uncertainties in the data and the theoretical prediction are treated as uncorrelated among all \mjj and \CHI bins, all systematic effects are treated as fully correlated among \mjj and \CHI bins.
The only exceptions are the systematic uncertainties related to response matrix modeling, detector simulation, and JER tails, whose correlations among \mjj bins are less well known; thus they are considered with additional independent nuisance parameters per \mjj bin to account for potential uncorrelated components.

\section{Search for BSM physics}

To search for signatures of BSM physics, the detector-level distributions
are compared to QCD and BSM predictions,
using a response matrix to obtain predictions at the detector level.
The predictions for each \mjj range are normalized to the number
of observed events in that range.
The log-likelihoods $L_\text{QCD}$ and $L_\text{QCD+BSM}$
for the QCD-only and QCD+BSM hypotheses, respectively, are defined
as products of Poissonian likelihood functions for each bin in the \CHI and \mjj regions.
At the detector level, the measured data counts can be treated as uncorrelated among all bins.
The asymptotic approximation~\cite{AsymptCLs,CMS-NOTE-2011-005} of the \CLs
criterion~\cite{junk:1999,read:2002}
is used to set exclusion limits on various BSM model parameters, using the CMS statistical analysis tool, \textsc{Combine}~\cite{CMS:2024onh}.
The limits obtained using this approximation were tested against the
\CLs limits obtained using ensembles of pseudo-experiments for several of the models
examined, and the differences were found to be negligible.
Nuisance
parameters are profiled with respect to the QCD-only and QCD+BSM models by maximizing the
corresponding likelihood functions.
The $p$-values for the two hypotheses,
$P_\text{QCD+BSM}(q > q_\text{obs})$ and $P_\text{QCD}(q < q_\text{obs})$,
are based on the log-likelihood ratio $q = -2 \ln (L_\text{QCD+BSM}/L_\text{QCD})$.
Limits on the QCD+BSM models are set based on the quantity
$\CLs = P_\text{QCD+BSM}(q > q_\text{obs}) /
(1-P_\text{QCD}(q < q_\text{obs}))$, which is required to be
less than 0.05 for a 95\% confidence level (\CL) exclusion.
Expected limits are obtained by using a QCD-only Asimov dataset satisfying the above requirement.

\begin{figure*}[htbp]
\centering
\includegraphics[width=0.45\textwidth]{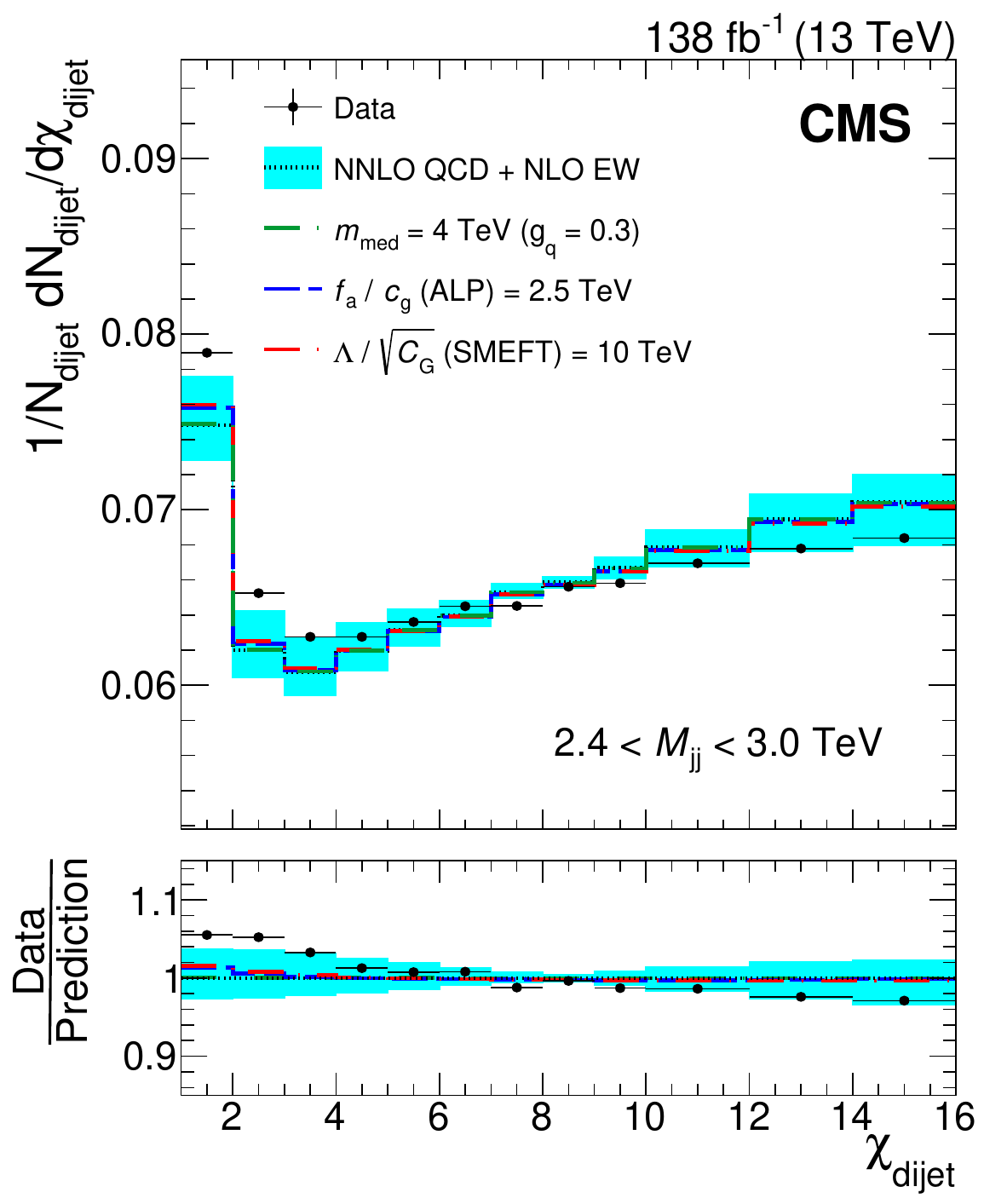}
\includegraphics[width=0.45\textwidth]{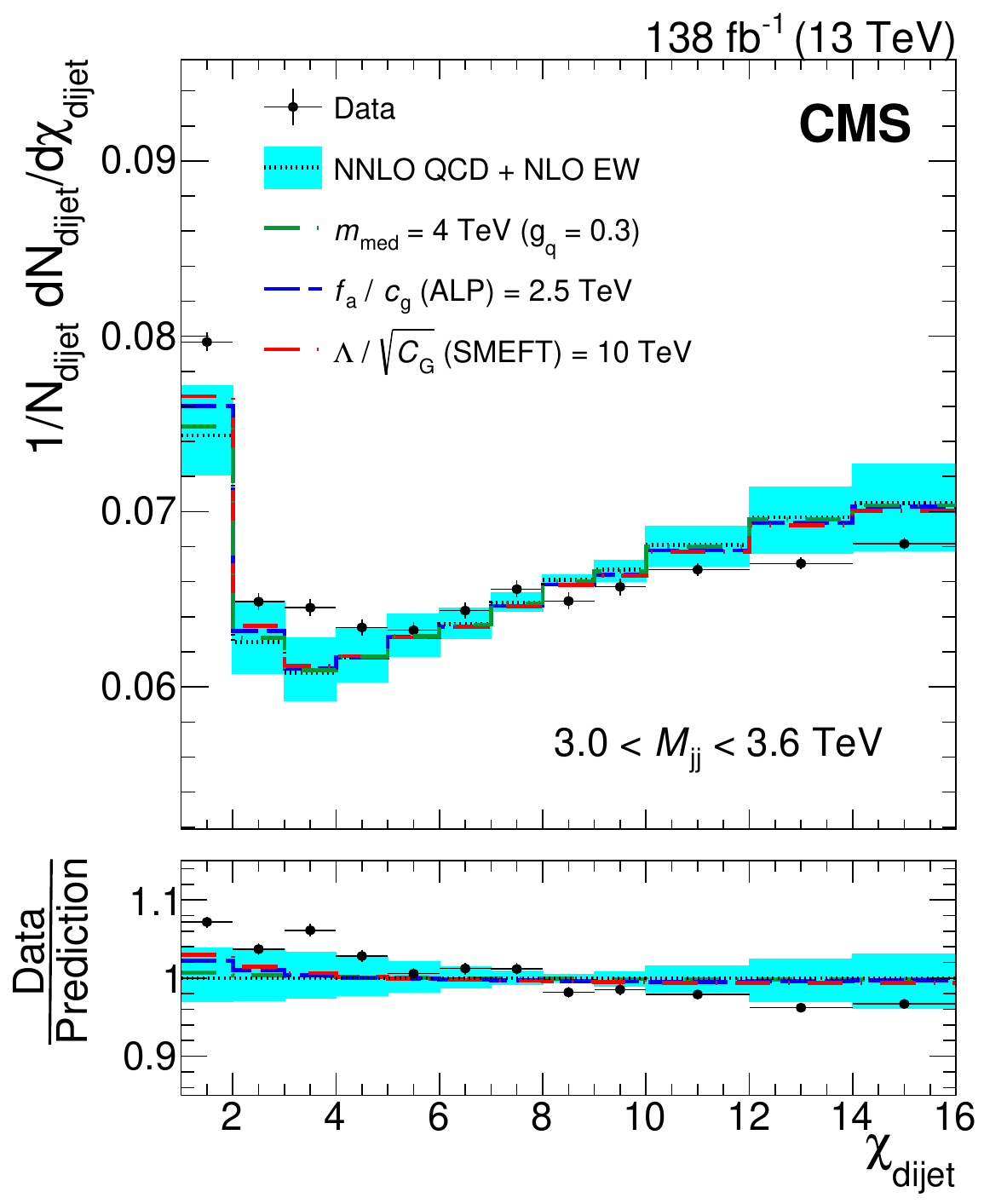} \\
\includegraphics[width=0.45\textwidth]{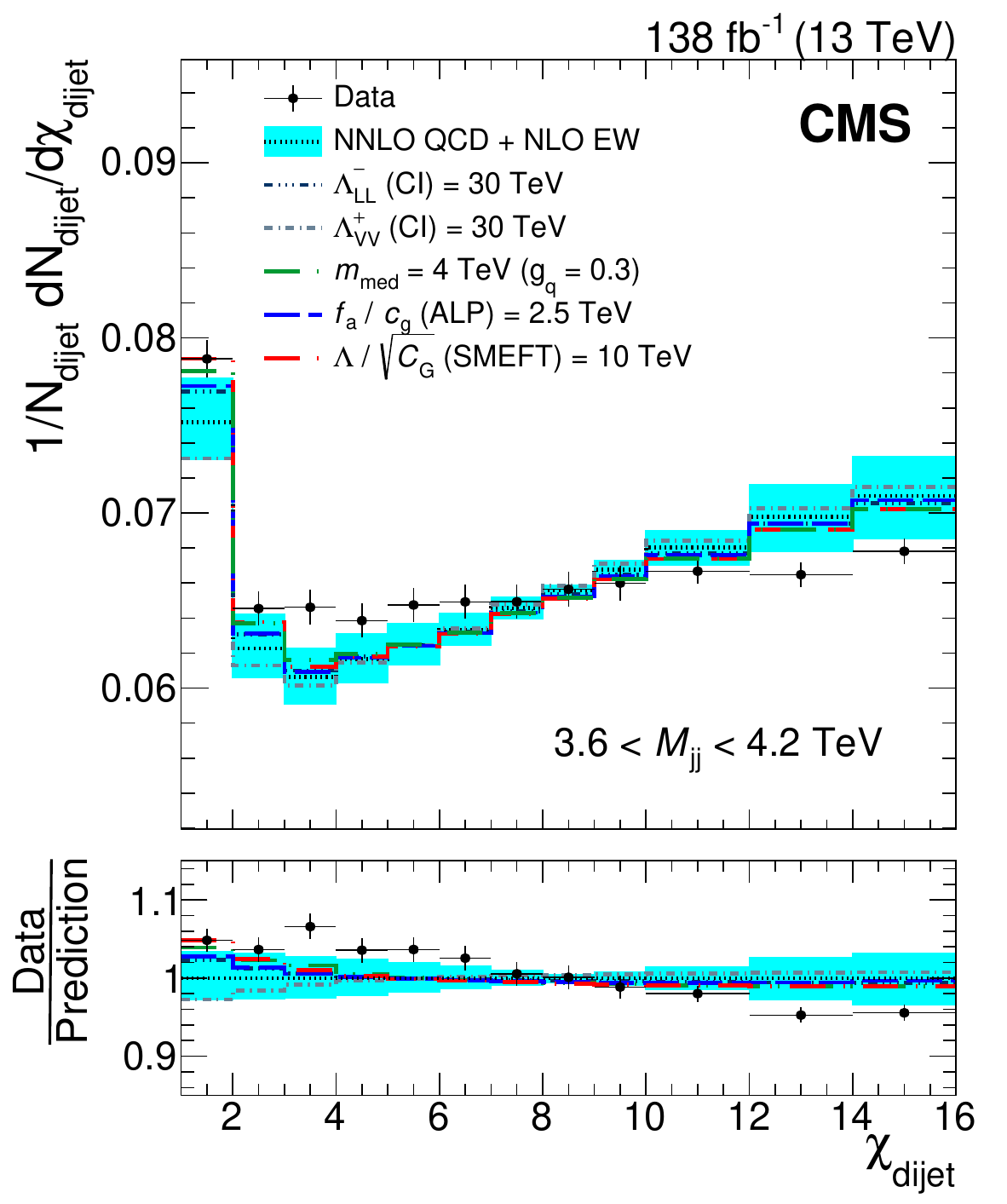}
\includegraphics[width=0.45\textwidth]{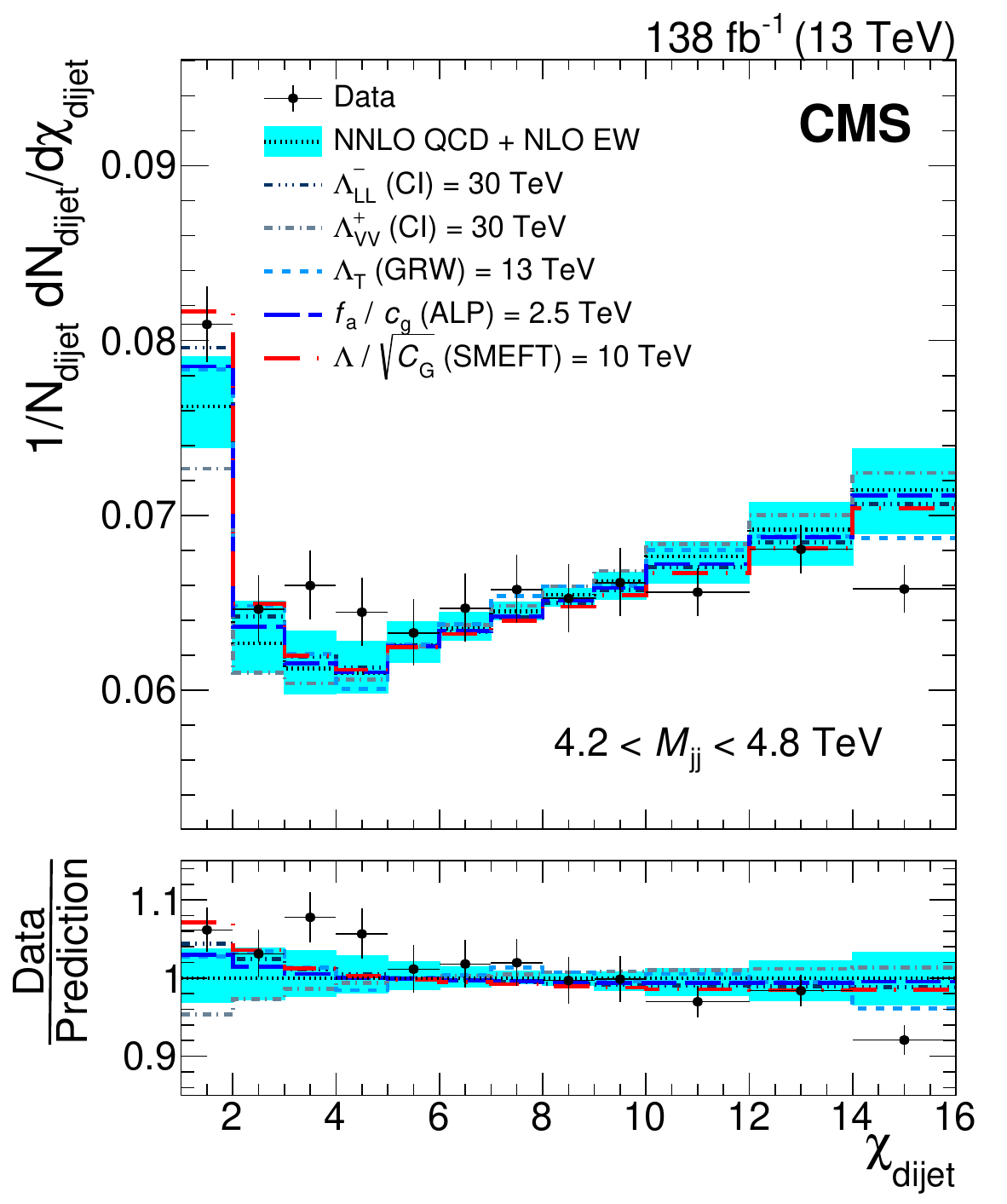}
\caption{Normalized \CHI distributions for the \mjj bins up to 4.8\TeV. The data distributions at the detector level (points) are compared to NNLO predictions, corrected for the detector response (black dotted lines).
 The vertical bars on the points represent statistical and experimental systematic uncertainties combined in quadrature.
 The horizontal bars show the bin widths.
 Theoretical uncertainties are indicated with the blue bands.
 The prediction from various BSM scenarios, with parameter values given in the legend, are shown by the different dot-dashed lines.
 The lower plots display the ratio of the data to the NNLO QCD + NLO EW predictions.
 }
\label{fig:data_results_detector1}
\end{figure*}

\begin{figure*}[htbp]
\centering
\includegraphics[width=0.45\textwidth]{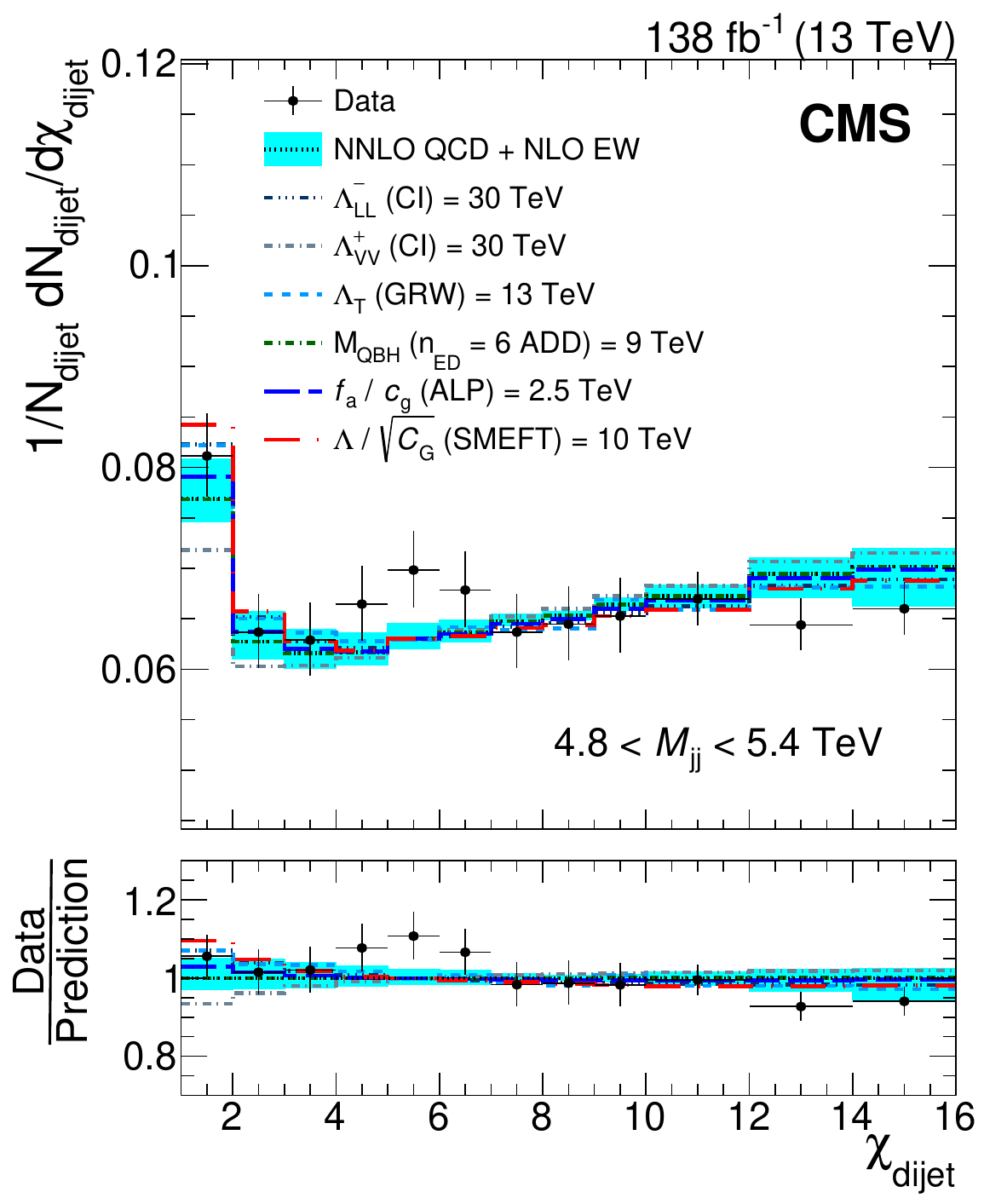}
\includegraphics[width=0.45\textwidth]{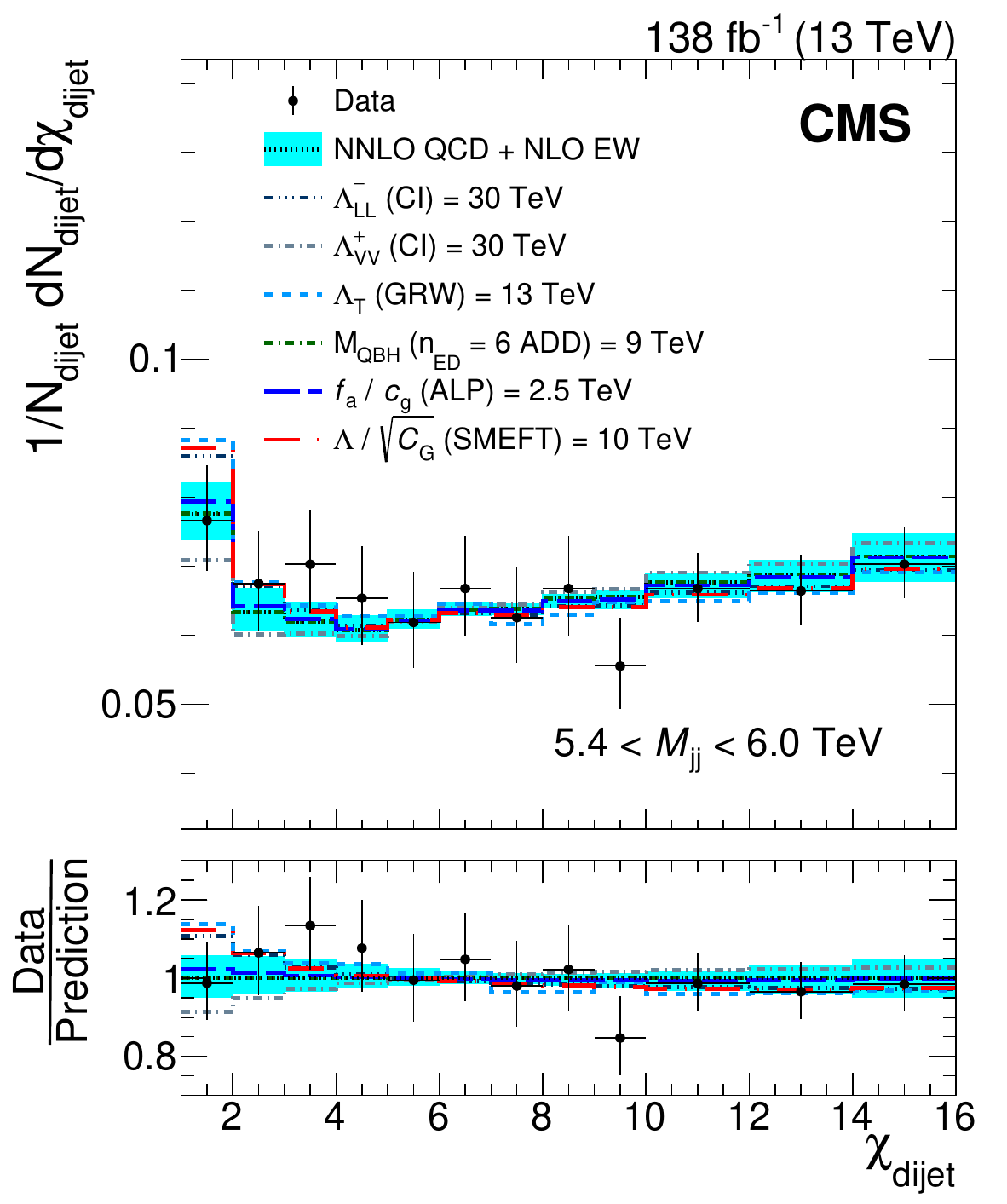} \\
\includegraphics[width=0.45\textwidth]{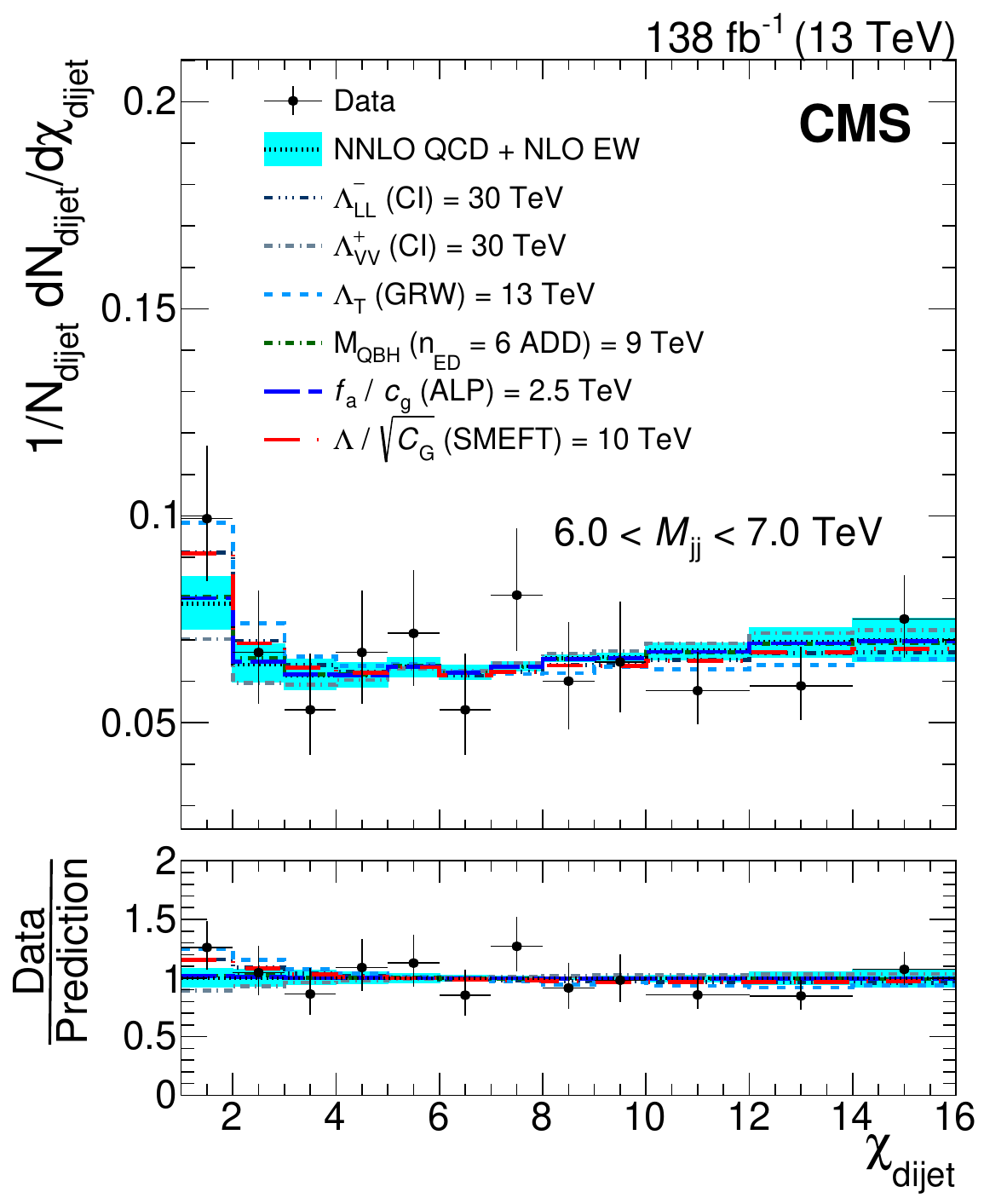}
\includegraphics[width=0.45\textwidth]{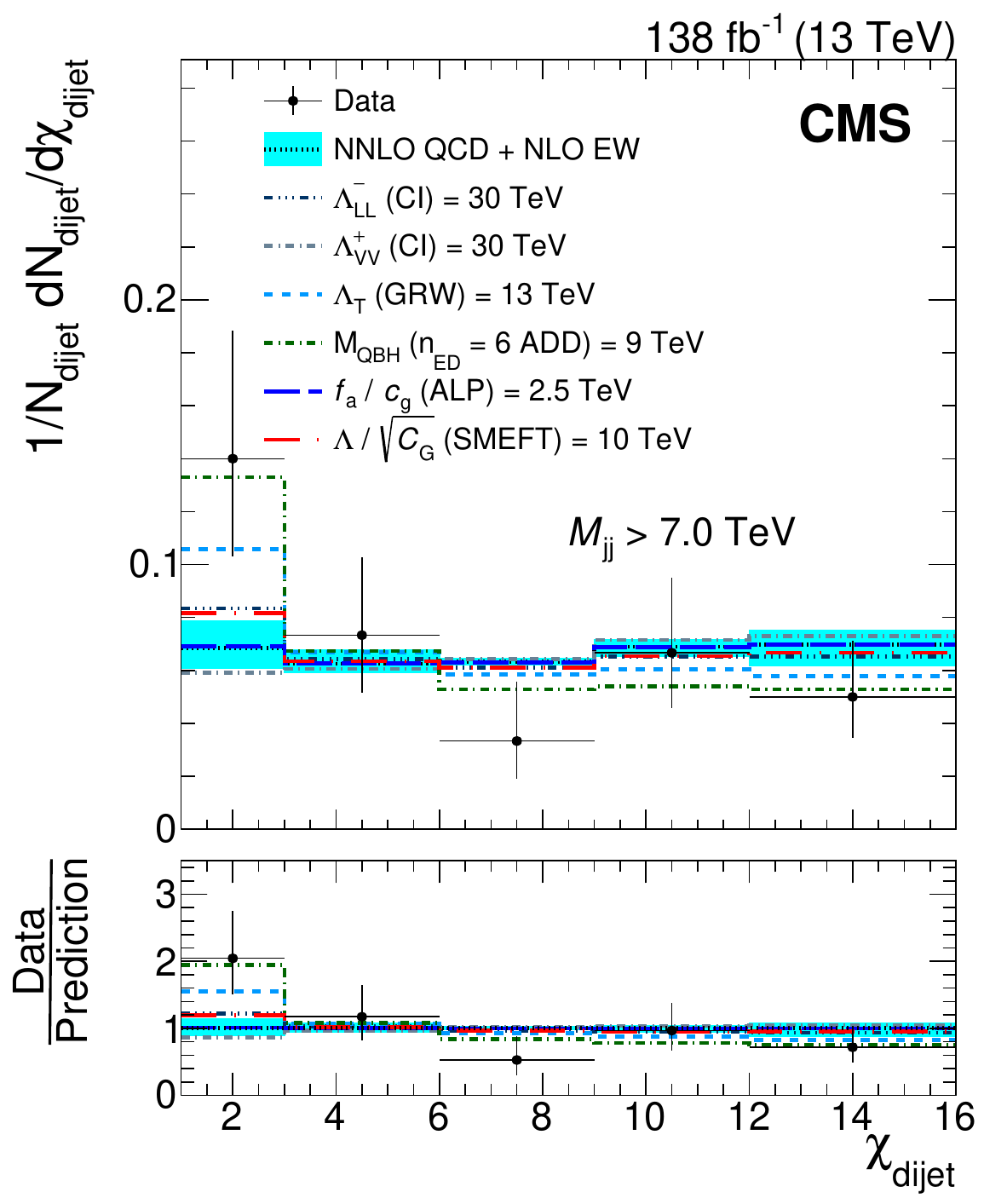}
\caption{Normalized \CHI distributions for the \mjj bins above 4.8\TeV. Notations as in Fig.~\ref{fig:data_results_detector1}.}
\label{fig:data_results_detector2}
\end{figure*}

In Figs.~\ref{fig:data_results_detector1}--\ref{fig:data_results_detector2}, the \CHI distributions at the detector level are compared with representative examples of BSM predictions, which account for detector effects by applying the response matrix.
The distributions are shown before the likelihood fit is carried out.
Due to the large number of data events in the 2.4--3.0\TeV \mjj bin, the likelihood fit strongly constrains the background nuisance parameters, such that the background shape agrees with the data after the fit (not shown here).
The signal distributions are only shown for a subset of \mjj ranges, since those bins dominate the sensitivity to the BSM signals that were considered.
While signatures of CI, ADD, and QBH mainly predict an excess at low \CHI in the highest \mjj bins, DM, ALP, and SMEFT predictions modify the \CHI distributions at lower \mjj as well.

In the \mjj bins below 4.8\TeV, the measured distributions show a small difference in shape compared to the QCD-only prediction by an amount similar to the experimental and theoretical uncertainties.
For \mjj in the range 4.8 to 7\TeV, generally good agreement is found between the measurement and the QCD-only prediction.
In the highest \mjj bin and the lowest \CHI bin, another small difference compared to the QCD prediction is seen.
The agreement of the data with QCD predictions is quantified by
calculating $P_\text{QCD}(q < q_\text{obs})$,
considering a CI signal in the likelihood ratio.
When taking into account all \mjj bins,
a significance of 0.3 standard deviations is found.
Considering individual mass bins alone,
the largest deviations are found in the 3--3.6\TeV and 7--13\TeV \mjj bins,
with a local significance of 2.0 and 2.4 standard deviations, respectively.
However, none of the signals considered would be localized
in a single mass bin; thus we find no evidence for a BSM signal.

The measured \CHI distributions are used to determine exclusion limits on the BSM models.
In the limit calculations, all \mjj ranges are included in the likelihoods.
To avoid sensitivity to statistical uncertainties in the signal predictions,
signal contributions are only accounted for in a subset of \mjj bins that contain a nonnegligible amount of signal, so that including those bins improves the expected limits by at least 1\%.
We use the mass bins with $\mjj>3.6\TeV$, 4.2\TeV, and 4.8\TeV in searching
for signals with the CI, ADD, and QBH models, respectively.
For the DM mediators, we use mass bins that cover $\mjj<1.2 m_\text{med}$.
For the ALP and SMEFT models, we only use the \mjj bins
below the BSM physics scale, $f_\text{a}$ and $\Lambda$, respectively.
The observed and expected exclusion limits on different models obtained
in this analysis at 95\% \CL are listed in Table~\ref{tab:limits}.

For ADD and QBH models, the observed limits are worse than the expected limits, due to a 
slight excess of events in the lowest \CHI bin for the highest $M_{\text{jj}}$ bin. 
The limit on CI $\Lambda_{\text{LL/RR}}^{+}$ is significantly lower than $\Lambda_{\text{LL/RR}}^{-}$, due to the destructive interference between the CI and QCD terms. 
The limit on $\Lambda_{\text{VV}}^{+}$ has two disjoint regions of observed exclusion. 
The lower limit on $\Lambda_{\text{VV}}^{+}$ is driven by pure CI terms and is followed by a gap in exclusion, 
when the CI and destructive interference terms balance each other. 
At large values of $\Lambda_{\text{VV}}^{+}$, the destructive interference dominates and leads to the additional exclusion region.
The most stringent lower limits to date are set on the scale of CIs,
graviton exchange, and QBHs.

\begin{table*}[htbp]
\centering
\topcaption{Observed and expected exclusion limits at 95\% \CL for various BSM models.
The $\pm1$ standard deviation values for the expected limits are quoted, as shown in the last column.
}
\label{tab:limits}
\begin{tabular}{lcc} 
Model & Observed limit & Expected limit\\
\hline
\noalign{\vskip 2.5pt}
CI $\Lambda_{\text{LL/RR}}^{+}$ (NLO) & 17\TeV & 18$^{+ 20}_{- 2}\TeV$ \\
CI $\Lambda_{\text{LL/RR}}^{-}$ (NLO) & 37\TeV & 34$^{+ 8}_{- 5}\TeV$ \\
CI $\Lambda_{\text{VV}}^{+}$ (NLO)    & $<$19 and 27--31\TeV & 41$^{+ 17}_{- 22}\TeV$ \\
CI $\Lambda_{\text{VV}}^{-}$ (NLO)    & 50\TeV & 45$^{+ 11}_{- 7}\TeV$ \\
CI $\Lambda_{\text{AA}}^{+}$ (NLO)    & 19\TeV& 39$^{+ 16}_{- 21}\TeV$ \\
CI $\Lambda_{\text{AA}}^{-}$ (NLO)    & 49\TeV& 44$^{+ 11}_{- 7}\TeV$ \\
CI $\Lambda_{\text{(V--A)}}^{+}$ (NLO) & 14.9\TeV& 14.0$^{+ 1.8}_{- 1.4}\TeV$ \\
CI $\Lambda_{\text{(V--A)}}^{-}$ (NLO) & 15.7\TeV & 14.6$^{+ 2.2}_{- 1.6}\TeV$ \\
ADD $\Lambda_{\text{T}}$ (GRW) & 13.4\TeV & 14.2 $\pm$ 1.0\TeV \\
ADD $M_{\text{S}}$ (HLZ) $n_{\text{ED}}=2$ & 15.0\TeV & 16.0 $\pm$ 1.1\TeV \\
ADD $M_{\text{S}}$ (HLZ) $n_{\text{ED}}=3$ & 15.9\TeV & 16.9 $\pm$ 1.2\TeV \\
ADD $M_{\text{S}}$ (HLZ) $n_{\text{ED}}=4$ & 13.4\TeV & 14.2 $\pm$ 1.0\TeV \\
ADD $M_{\text{S}}$ (HLZ) $n_{\text{ED}}=5$ & 12.1\TeV & 12.8 $\pm$ 0.9\TeV \\
ADD $M_{\text{S}}$ (HLZ) $n_{\text{ED}}=6$ & 11.3\TeV& 11.9 $\pm$ 0.8\TeV \\
QBH $M_{\text{QBH}}$ (ADD, $n_{\text{ED}}=6$) & 8.5\TeV & 9.1 $\pm$ 0.3\TeV \\
QBH $M_{\text{QBH}}$ (RS1)  & 6.3\TeV  & 6.8 $\pm$ 0.4\TeV \\ 
DM mediator $m_{\text{med}}$ & 6.2\TeV & 6.2 $\pm$ 0.8\TeV \\ 
ALP $c_{\text{g}}/f_{\text{a}}$  & $0.42\TeV^{-1}$ & 0.38 $\pm$ $0.03\TeV^{-1}$ \\ 
SMEFT $C_{\text{G}}/\Lambda^2$  & $0.0076\TeV^{-2}$ & 0.0081 $\pm$ $0.0012\TeV^{-2}$ \\ 
\end{tabular}
\end{table*}

\begin{figure}[htbp]
\centering
\includegraphics[width=0.45\textwidth]{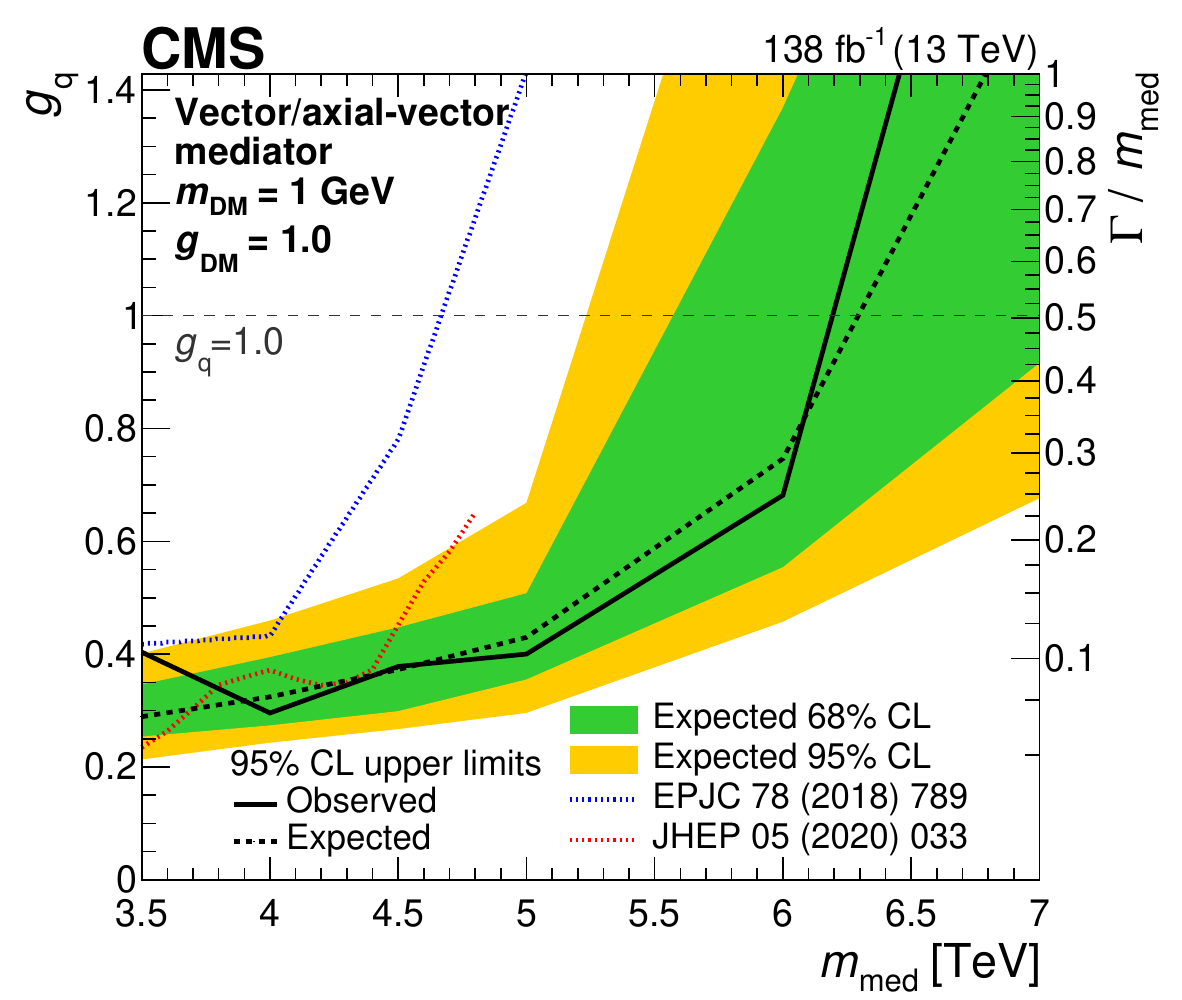}
\caption{The 95\% \CL upper limits on the universal quark coupling $g_{\text{q}}$ as functions of a vector or axial-vector mediator mass, with $g_{\text{DM}}=1.0$ and $m_{\text{DM}}=1\GeV$.
The observed (solid line) and expected (dashed line) limits and their 68\% and 95\% confidence intervals (shaded bands) are shown.
A dashed horizontal line shows the coupling strength for a benchmark DM mediator with $g_{\text{q}}=1$.
Results are compared to the previous searches using dijet angular~\cite{CMS:2018ucw} (blue dotted line) and dijet mass~\cite{CMS:2019gwf} (red dotted line) distributions.
The vertical scale on the right axis gives the corresponding values of $\Gamma/m_{\text{med}}$.
}
\label{fig:DMcoupling}
\end{figure}

For a vector or axial-vector DM mediator with $g_{\text{DM}}=1.0$ and $m_{\text{DM}}=1\GeV$, we compute limits as functions of the mediator mass, which are shown in Fig.~\ref{fig:DMcoupling}.
Unlike our previous search from dijet angular distributions~\cite{CMS:2018ucw} using only data with $0.5 m_\text{med}<\mjj<1.2 m_\text{med}$, we consider all mass bins satisfying $\mjj<1.2 m_\text{Med}$ to better constrain the background uncertainties.
Compared to previous searches for dijet resonances~\cite{ATLAS:2019fgd,CMS:2019gwf}, this search sets the most stringent limits on DM mediator masses between 4.5 and 6.5\TeV, where the mediator width is 10\% of its mass or larger.

\begin{figure}[htbp!]
\centering
\includegraphics[width=0.45\textwidth]{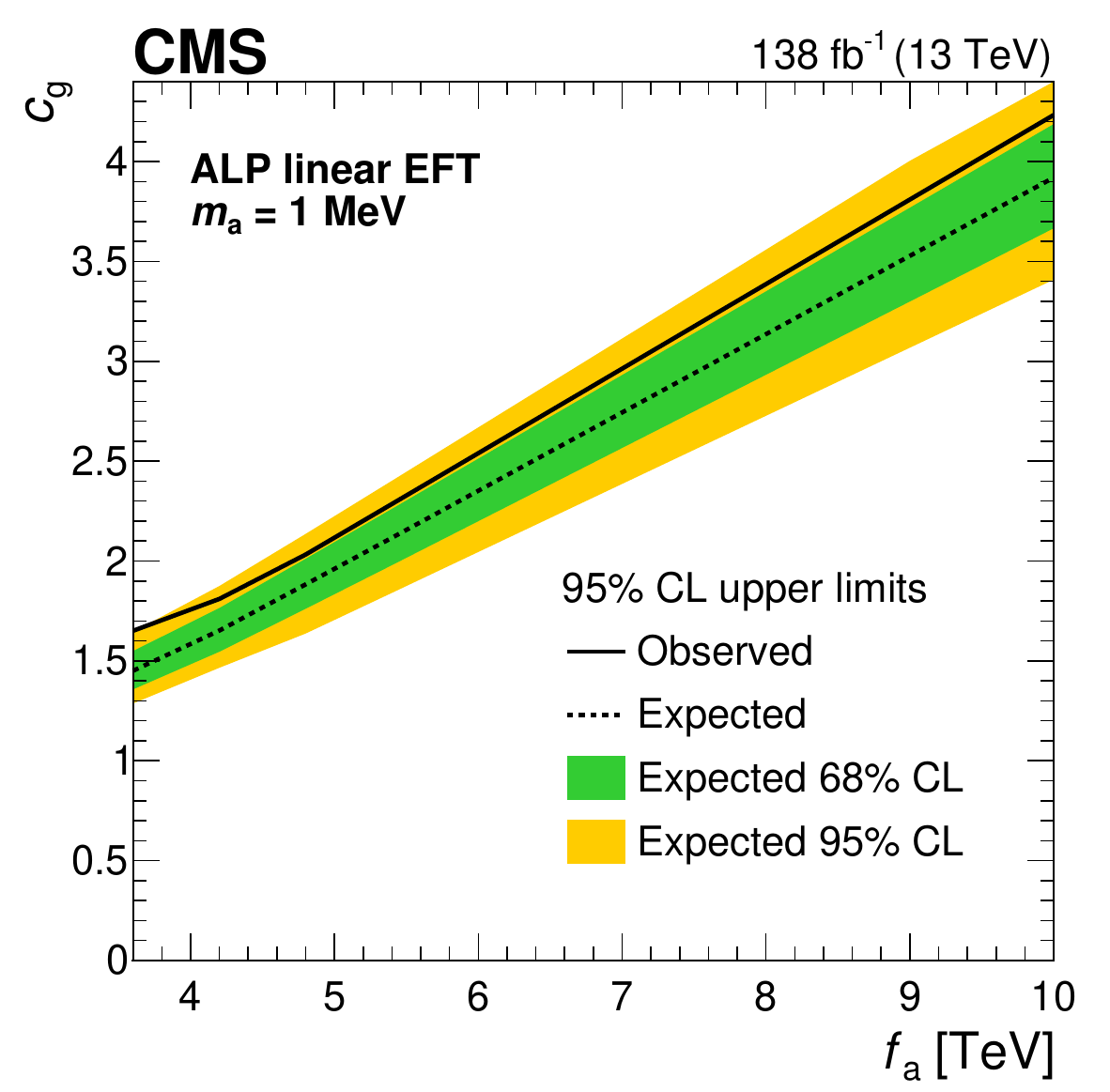}
\includegraphics[width=0.45\textwidth]{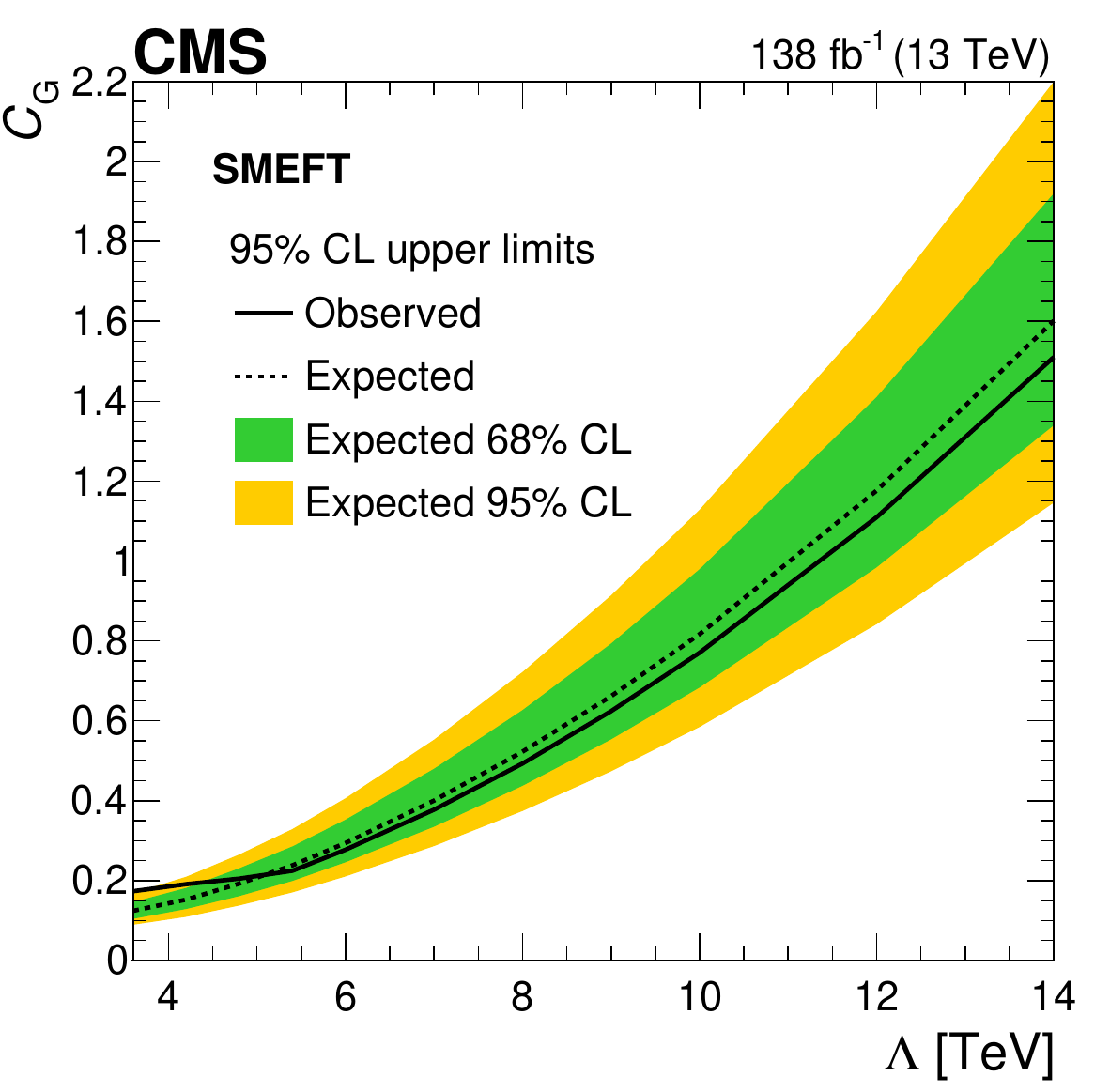}
\caption{The 95\% \CL upper limits on the ALP gluon coupling (\cmsLeft), $c_\text{g}$, in an EFT with a linear realization of electroweak symmetry breaking as functions of the characteristic energy scale, $f_\text{a}$, assuming $m_\text{a}=1\MeV$. Only \CHI distributions with $\mjj<f_\text{a}$ are used to obtain the limits on $c_\text{g}$.
The 95\% \CL upper limits on the anomalous triple-gluon coupling (\cmsRight), $C_\text{G}$, in SMEFT as functions of the BSM physics energy scale, $\Lambda$. Only \CHI distributions with $\mjj<\Lambda$ are used to obtain the limits on $C_\text{G}$. The observed limits (solid lines), expected limits (dashed lines), and their 68\% and 95\% confidence intervals (shaded bands) are shown.
}
\label{fig:EFTcouplings}
\end{figure}

In Fig.~\ref{fig:EFTcouplings} (\cmsLeft), we present limits on the ALP gluon coupling, $c_\text{g}$, as a function of the characteristic energy scale, $f_\text{a}$.
The limit on the ratio $f_\text{a}/c_\text{g}$ for energy scales $f_\text{a}$ beyond the reach of this measurement can be inferred from any point on the limit curve, which is linear above 7\TeV (our highest \mjj bin).
While an earlier reinterpretation using simplified assumptions reported an exclusion bound at $c_{\text{g}}/f_{\text{a}}=0.33\TeV^{-1}$~\cite{Gavela:2019cmq}, our limit is slightly weaker at $c_{\text{g}}/f_{\text{a}}=0.42\TeV^{-1}$.
An important difference between the previous limit and our result is that we consider 2- and 3-jet production in \MGvATNLO for signal and interference, while the previous result only considered $2\to2$ processes for the signal, interference, and background.
The 3-jet process spreads out the signal in the \CHI variable.

In Fig.~\ref{fig:EFTcouplings} (\cmsRight), the 95\% \CL upper limits on the anomalous triple-gluon coupling, $C_\text{G}$, in SMEFT is shown as a function of the BSM physics energy scale, $\Lambda$.
While a previous result using simplified assumptions on the experimental analysis reported a 95\% \CL upper limit on $C_{\text{G}}/\Lambda^2$ of $0.032\TeV^{-2}$~\cite{Goldouzian:2020wdq}, our 95\% \CL upper limit of $C_{\text{G}}/\Lambda^2=0.0076\TeV^{-2}$ is stronger.
It is also more stringent than the upper limit obtained from multijet production of $0.036\TeV^{-2}$~\cite{Krauss:2016ely},
as well as from top quark production~\cite{Ellis:2020unq}.
It should be noted that top quark production is uniquely sensitive to linear contributions of $C_\text{G}/\Lambda^2$, while dijet production is dominated by quadratic contributions of order $C_\text{G}^2/\Lambda^4$.

\section{Unfolding for the detector effects}
\label{sec:unfolding}

In the unfolding procedure used in this analysis, a measured distribution at the detector level is corrected for detector effects, yielding a particle-level distribution.
In a maximum likelihood template fit, the detector response matrix is inverted to obtain the measured distribution at the particle level.
The approach is similar to the one outlined in Ref.~\cite{S_Schmitt_2012}, though a maximum likelihood fit instead of a least-squares fit is used.
The signal strength modifiers $\mu_j$ are the ratios of the number of events predicted by the 
fit to data to the expected number of events at the particle level in simulation. The total number of 
expected events $y_i$ in bin $i$ at the detector level is given by 
\begin{equation}
    y_i = \sum_{j=1}^M R_{ij} \mu_j  x_j + b_i.
\end{equation}
Here, $R_{ij}$ is the response matrix corresponding to the number of events in a bin $i$ at the detector level, given the number of events in bin $j$ at the particle level, $M$ is the total number of bins in the particle-level distribution, and $b_i$ corresponds to the total number of expected background events in the $i$-th bin at the detector level.
In our measurement, the background is negligible.
If the response matrix can be inverted, the expected number of true events $x_j\mu_j$ at the particle level is given by:
\begin{equation}
    x_j\mu_j = \sum_{i=1}^N R^{-1}_{ji} (y_i - b_i),\label{eq:matrix_inversion}
\end{equation}
where $N$ is the total number of bins in the detector-level distribution.
This solution is obtained by minimizing the negative log-likelihood function:
\begin{equation}
    - 2 \ln \mathcal{L}(\vec{\mu}) = - 2 \sum_i^N \ln P(n_i, y_i),\label{eq:likelihood_part1}
\end{equation}
where $n_i$ is the total number of data events in the $i$-th detector-level bin (for all \mjj and \CHI bins), $P(n_i, y_i)$ is the Poisson probability distribution, and no regularization term is added.
Terms accounting for systematic uncertainties have been added as nuisance parameters to the negative log-likelihood function and profiled.
The dijet angular distributions at the particle level are measured in bins with $\mjj>3.0\TeV$, accounting for migrations from the 2.4--3.0\TeV bin at the detector level into the higher-mass \mjj bins at the particle level.
The unfolding changes the shape of the \CHI distributions by $<$1\% and $<$10\% for all values of \CHI in the lowest and highest \mjj ranges, respectively. 

\begin{figure*}[htbp!]
\centering
\includegraphics[width=0.45\textwidth]{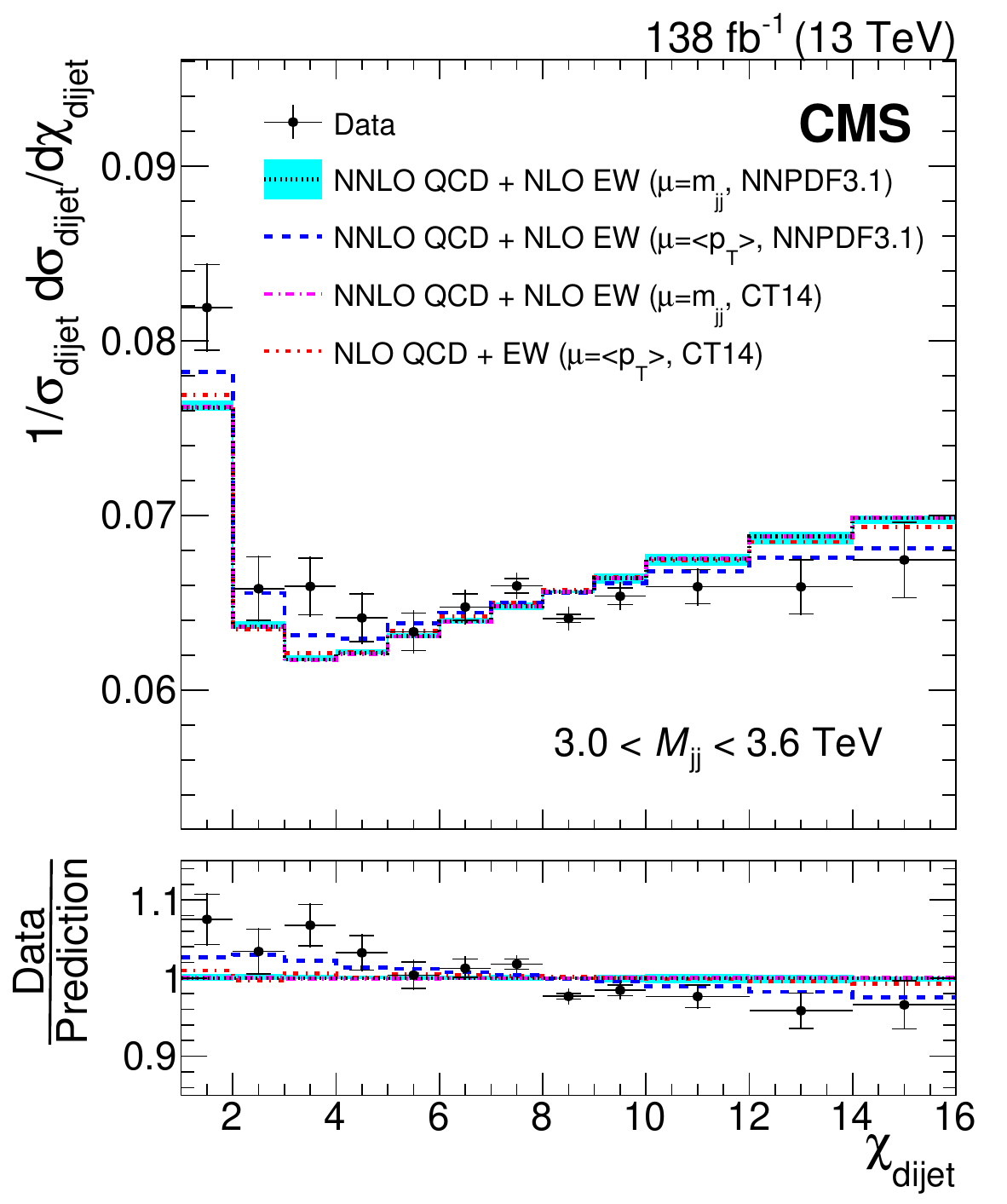}
\includegraphics[width=0.45\textwidth]{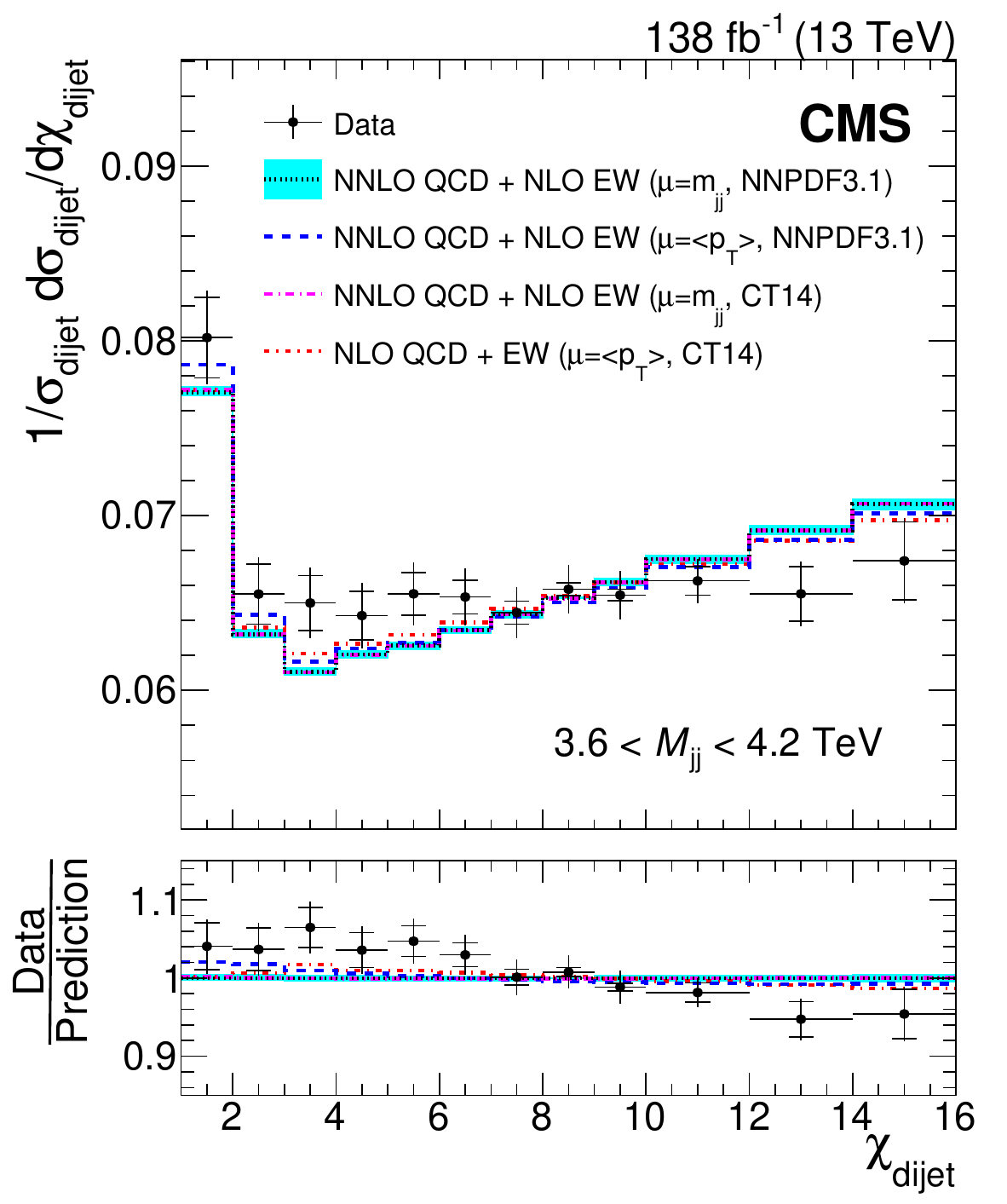} \\
\includegraphics[width=0.45\textwidth]{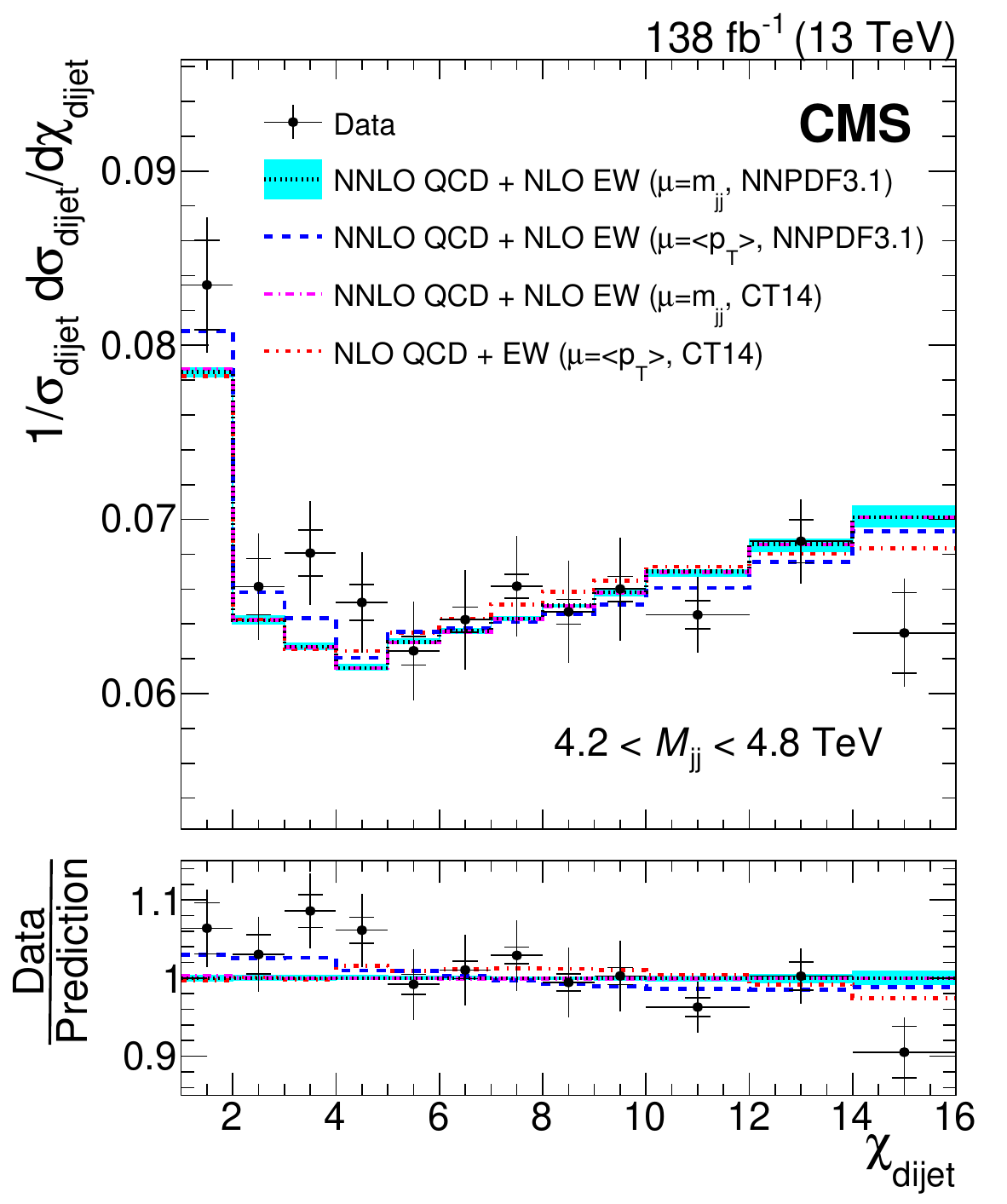}
\includegraphics[width=0.45\textwidth]{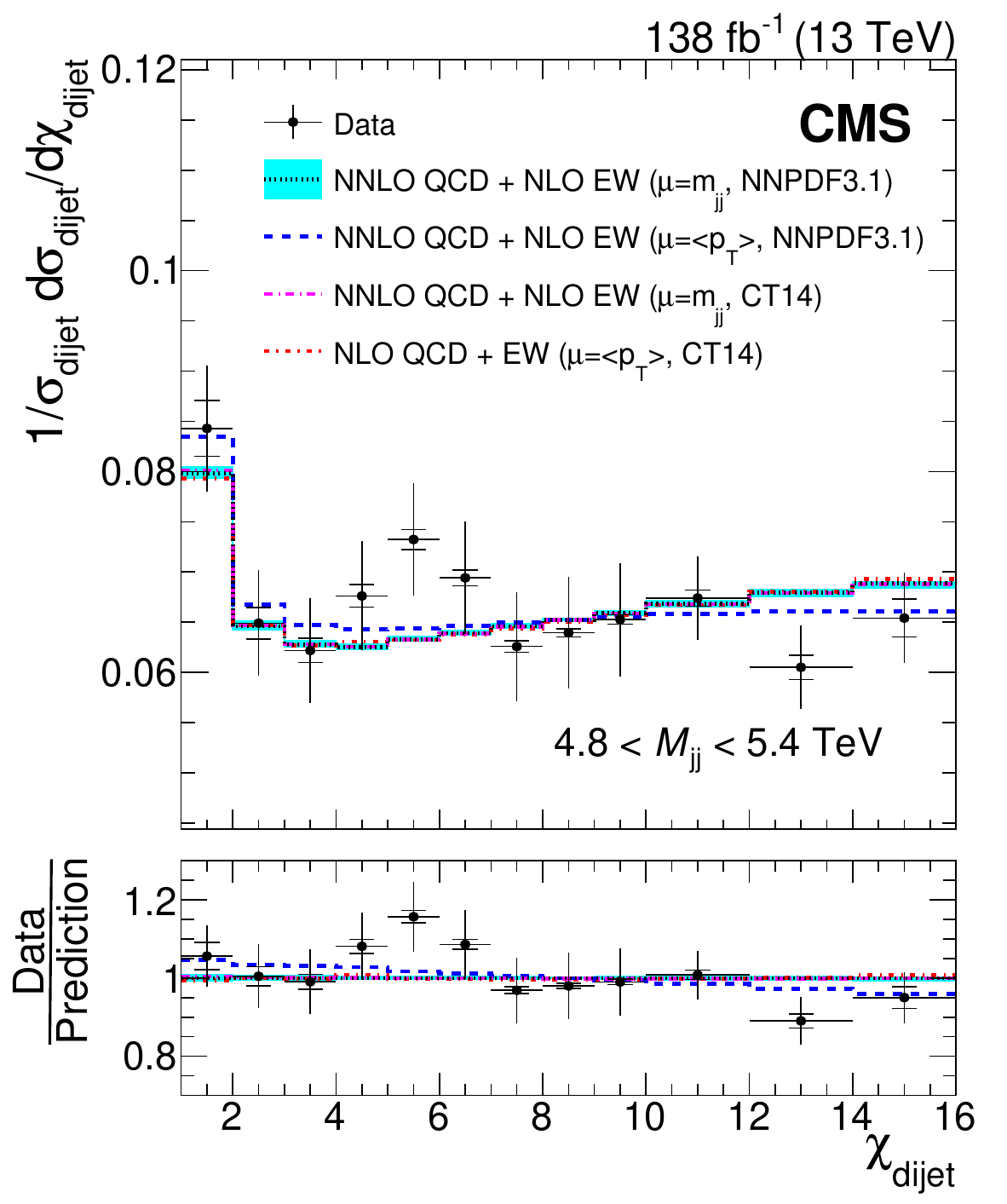}
\caption{Normalized \CHI distributions from data (points) corrected for detector effects for four different ranges of \mjj.
The inner vertical bars on the points correspond to the systematic uncertainties in the data, and the outer ones to the total uncertainties.
The horizontal bars give the bin widths.
Theoretical NNLO QCD + NLO EW predictions are shown for central scale choices $\mu_{\mathrm{F}} = \mu_{\mathrm{R}} = \mjj$ (black dotted line) and $\mu_{\mathrm{F}} = \mu_{\mathrm{R}} = \ptave$ (blue dashed line), both using the NNPDF3.1 PDF set.
The prediction using the alternative CT14 PDF set (purple long-dashed-dotted line), and the NLO predictions from a previous CMS publication~\cite{CMS:2021yzl} (red short-dashed-dotted lines) are also plotted.
 Theoretical uncertainties with the central scale $\mu_{\mathrm{F}} = \mu_{\mathrm{R}} = \mjj$ and using NNPDF3.1 are indicated with the gray bands.
The lower plots display the ratio of the data to the NNLO QCD + NLO EW predictions as well as the ratios of alternative predictions to the default prediction.
 }
\label{fig:data_results_unfolded1}
\end{figure*}

\begin{figure*}[htbp!]
\centering
\includegraphics[width=0.45\textwidth]{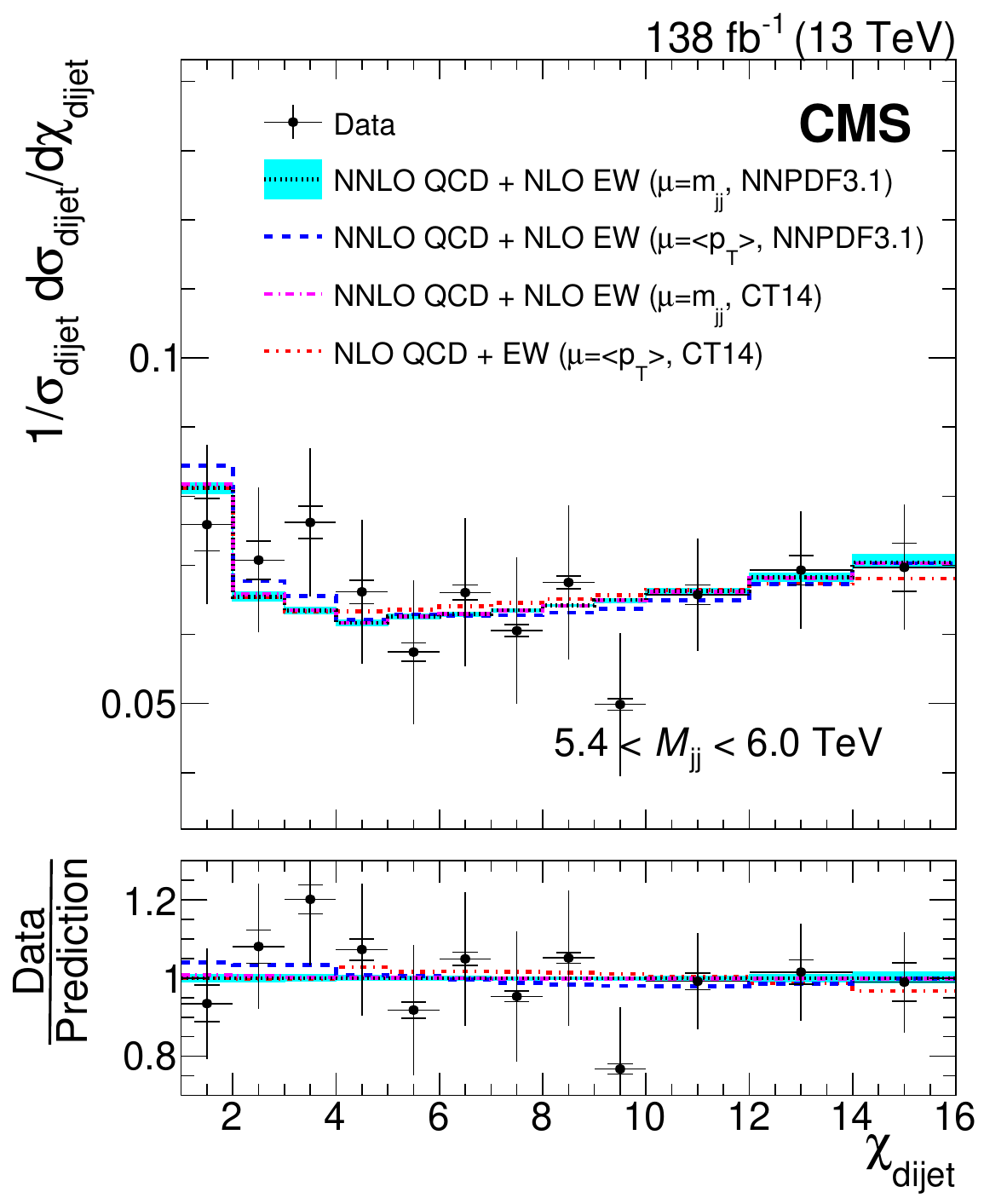} 
\includegraphics[width=0.45\textwidth]{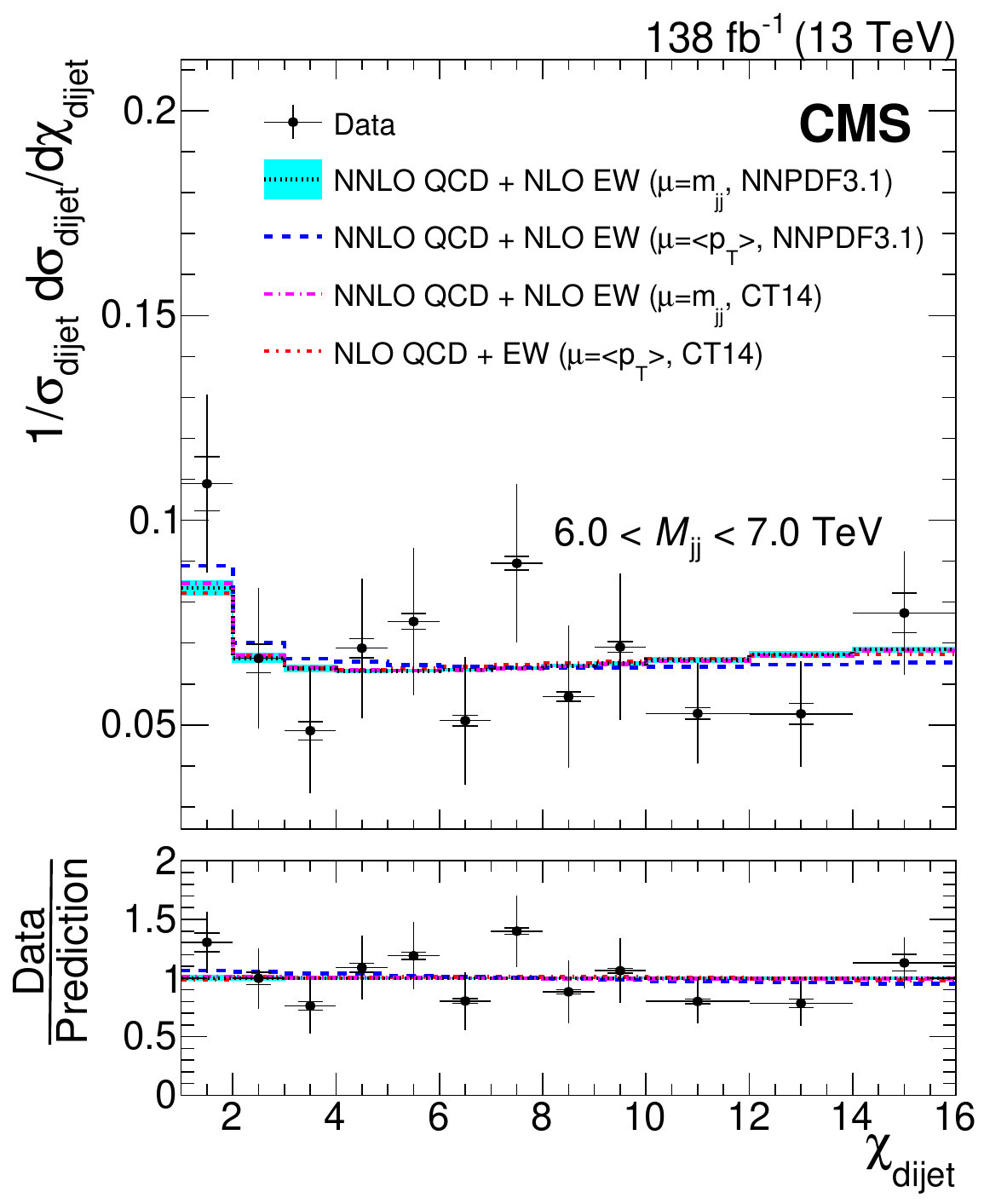} \\
\includegraphics[width=0.45\textwidth]{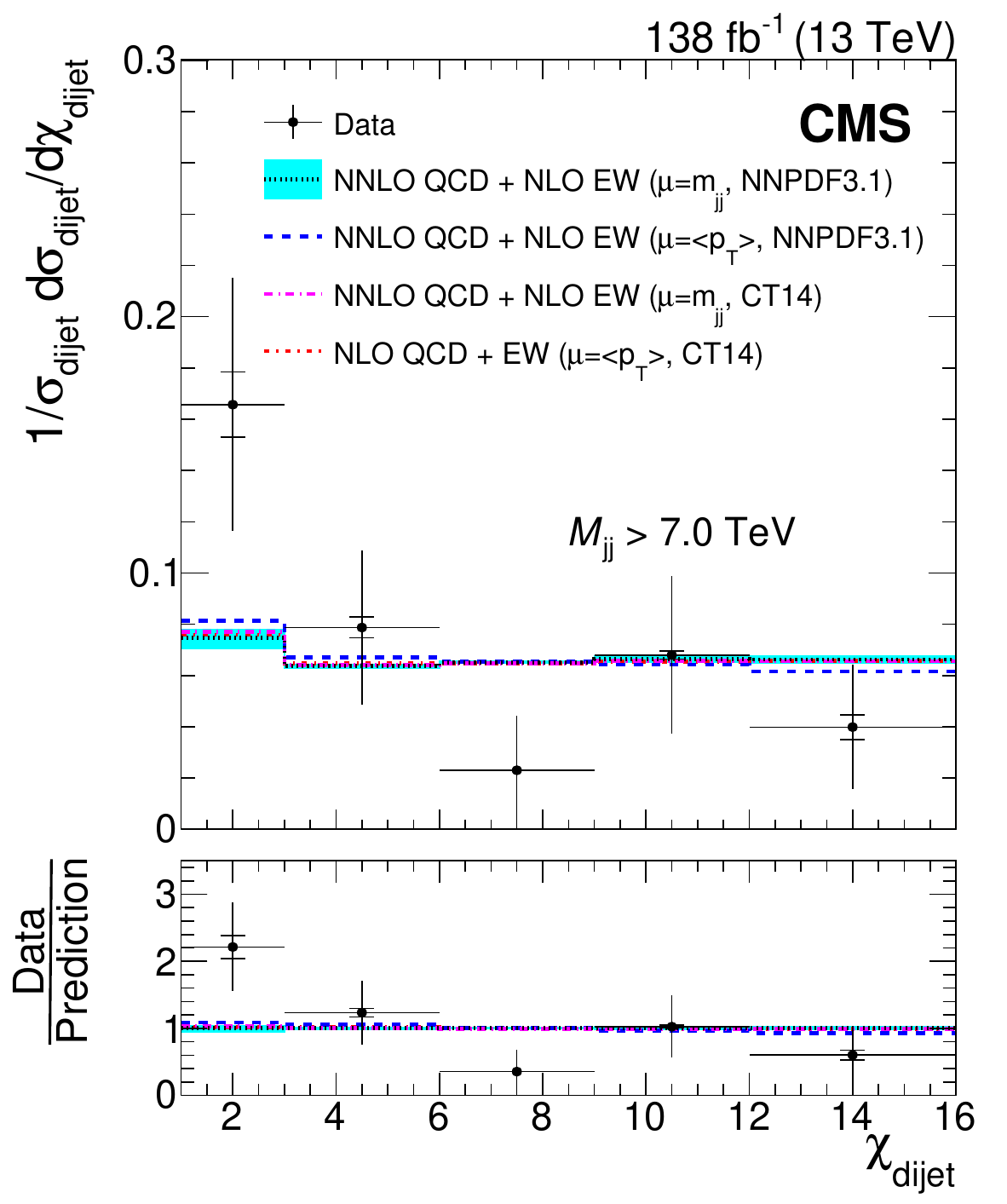}
\caption{Normalized \CHI distributions corrected for detector effects for three different ranges of \mjj. Notations as in Fig.~\ref{fig:data_results_unfolded1}.
 }
\label{fig:data_results_unfolded2}
\end{figure*}

\begin{figure}[htbp!]
\centering
\includegraphics[width=0.5\textwidth]{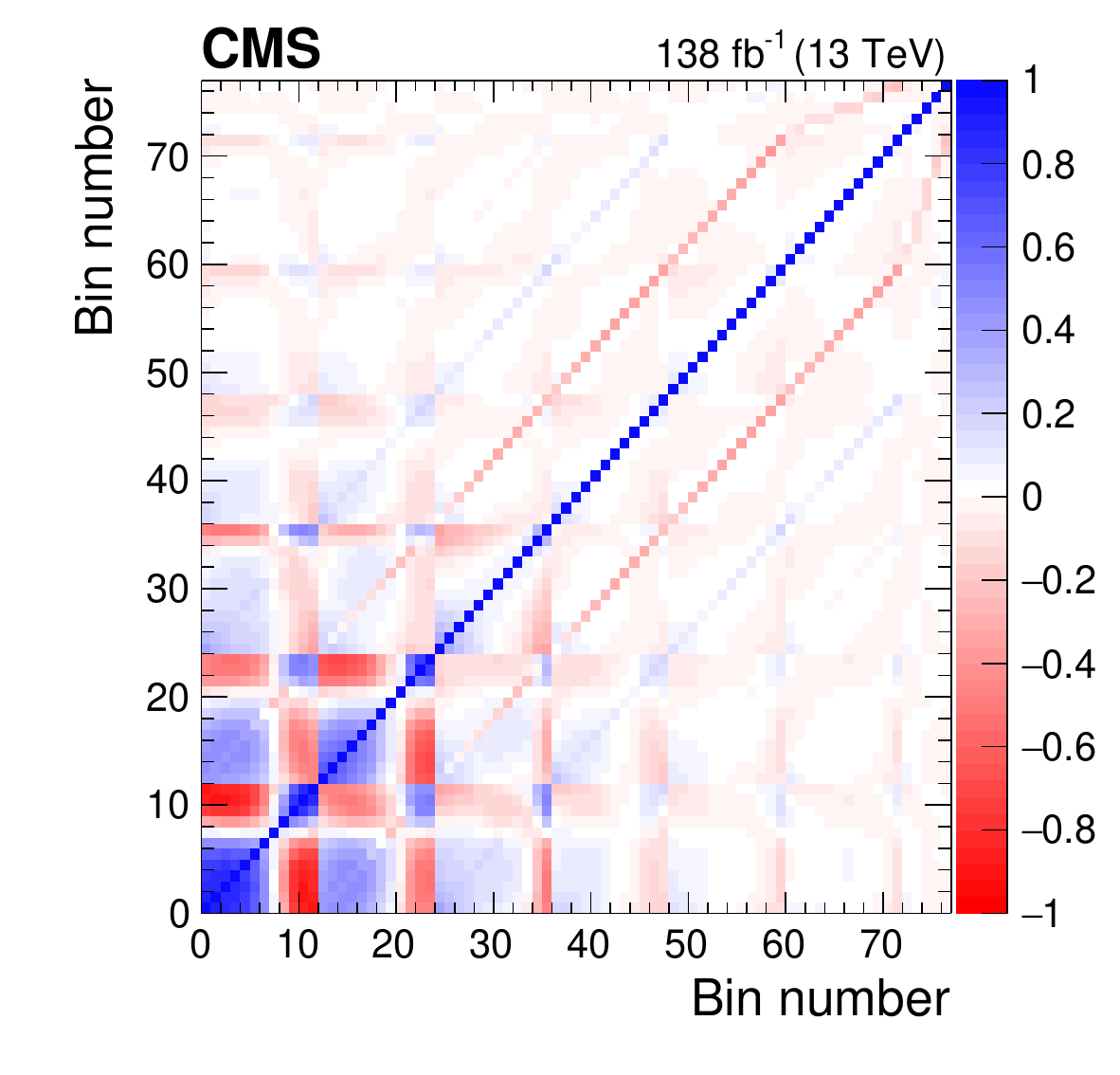}
\caption{The correlation matrix of the maximum likelihood estimators of the signal strength modifiers. The matrix is obtained after the fit to data. The bin numbers correspond to 11 times the index (starting at 0) of the \mjj bin (3.0, 3.6, 4.2, 4.8, 5.4, 6.0, 7.0) \GeV plus the index (starting at 0) of the \CHI bin (1, 2, 3, 4, 5, 6, 7, 8, 9, 10, 12, 14).
}
\label{fig:correlation_matrix}
\end{figure}

The resulting particle-level dijet angular distributions are shown in Figs.~\ref{fig:data_results_unfolded1}--\ref{fig:data_results_unfolded2}.
The correlations between bins introduced by the unfolding are provided in Fig.~\ref{fig:correlation_matrix}.
For $\mjj = 4.8$--$7.0\TeV$, generally good agreement is found between the measurement
and the NNLO predictions with both choices of the central scale.
In the \mjj bins below 4.8\TeV, the two NNLO NNPDF3.1 predictions disagree by an amount larger than the uncertainty in the prediction with the central scale $\mu_{\mathrm{F}} = \mu_{\mathrm{R}} = \mjj$.
We observe a small tension between the data and the NNLO predictions when considering only the six variations of scales around their central value, but not when using the alternative scale choice of \ptave.
An alternative prediction using the CT14 PDF set at NNLO shows negligible differences with respect to the default prediction using NNPDF3.1, demonstrating the insensitivity of the observable to the choice of PDF.
For reference, we also show the NLO prediction with the CT14 PDFs and \ptave scale used in the previous publication~\cite{CMS:2021yzl}, which also agrees well with the default prediction.

\ifthenelse{\boolean{cms@external}}{\clearpage}{}
\section{Summary}

Normalized dijet angular distributions have been measured
with the CMS detector over a wide range of dijet invariant masses,
using proton-proton collisions at the LHC, corresponding to an integrated luminosity of 138\fbinv.
For the first time, the dijet angular distributions corrected for detector effects have been compared with predictions of perturbative quantum chromodynamics at next-to-next-to-leading order.
Predictions assuming two different choices for the factorization and renormalization scales, $\mu_{\mathrm{F}} = \mu_{\mathrm{R}} = \mjj$ or \ptave, are found to differ by more than the six variations of scales commonly used to quantify the uncertainty in the predictions.
While data are generally found to be in agreement with the standard model predictions, a small difference in the shape of the normalized distributions is seen for dijet masses below 4.8\TeV and above 6.0\TeV.

The dijet angular distributions at the detector level are used to set lower limits on beyond the standard model theories, including quark contact interaction models with next-to-leading-order quantum chromodynamic corrections, extra dimension models, quantum black hole production, dark-matter mediators, axion-like particles, and standard model effective field theory.
The most stringent lower limits to date are set on the scale of graviton exchange.
In the Giudice--Rattazzi--Wells convention, virtual graviton exchange is excluded up to a scale of 13.4\TeV at 95\% confidence level (\CL).
The production of quantum black holes is excluded for masses below 8.5 and 6.3\TeV at 95\% \CL, depending on the scenario.

Limits for different contact interaction models are obtained.
In the benchmark scenario, valid to next-to-leading order in quantum chromodynamics and in which only left-handed quarks participate, quark contact interactions are excluded to a scale of 17 (37)\TeV at 95\% \CL for destructive (constructive) interference.
Mediators in a simplified model of interactions between quarks and dark matter with masses between 4 and 6.2\TeV are excluded at 95\% \CL for vector and axial-vector mediators.
The ratio of the gluon operator coefficient $c_{\text{g}}$ to the characteristic high-energy scale $f_{\text{a}}$, $c_{\text{g}}/f_{\text{a}}$, is determined to have a 95\% \CL upper limit of $0.42\TeV^{-1}$.
The anomalous triple-gluon coupling, $C_{\text{G}}/\Lambda^2$, in a standard model effective field theory is constrained to be lower than $0.0076\TeV^{-2}$ at 95\% \CL.

\begin{acknowledgments}
We congratulate our colleagues in the CERN accelerator departments for the excellent performance of the LHC and thank the technical and administrative staffs at CERN and at other CMS institutes for their contributions to the success of the CMS effort. In addition, we gratefully acknowledge the computing centers and personnel of the Worldwide LHC Computing Grid and other centers for delivering so effectively the computing infrastructure essential to our analyses. Finally, we acknowledge the enduring support for the construction and operation of the LHC, the CMS detector, and the supporting computing infrastructure provided by the following funding agencies: SC (Armenia), BMBWF and FWF (Austria); FNRS and FWO (Belgium); CNPq, CAPES, FAPERJ, FAPERGS, and FAPESP (Brazil); MES and BNSF (Bulgaria); CERN; CAS, MoST, and NSFC (China); MINCIENCIAS (Colombia); MSES and CSF (Croatia); RIF (Cyprus); SENESCYT (Ecuador); ERC PRG and PSG, TARISTU24-TK10 and MoER TK202 (Estonia); Academy of Finland, MEC, and HIP (Finland); CEA and CNRS/IN2P3 (France); SRNSF (Georgia); BMFTR, DFG, and HGF (Germany); GSRI (Greece); NKFIH (Hungary); DAE and DST (India); IPM (Iran); SFI (Ireland); INFN (Italy); MSIT and NRF (Republic of Korea); MES (Latvia); LMTLT (Lithuania); MOE and UM (Malaysia); BUAP, CINVESTAV, CONACYT, LNS, SEP, and UASLP-FAI (Mexico); MOS (Montenegro); MBIE (New Zealand); PAEC (Pakistan); MES, NSC, and NAWA (Poland); FCT (Portugal); MESTD (Serbia); MICIU/AEI and PCTI (Spain); MOSTR (Sri Lanka); Swiss Funding Agencies (Switzerland); MST (Taipei); MHESI (Thailand); TUBITAK and TENMAK (T\"{u}rkiye); NASU (Ukraine); STFC (United Kingdom); DOE and NSF (USA).

\hyphenation{Rachada-pisek} Individuals have received support from the Marie-Curie program and the European Research Council and Horizon 2020 Grant, contract Nos.\ 675440, 724704, 752730, 758316, 765710, 824093, 101115353, 101002207, 101001205, and COST Action CA16108 (European Union); the Leventis Foundation; the Alfred P.\ Sloan Foundation; the Alexander von Humboldt Foundation; the Science Committee, project no. 22rl-037 (Armenia); the Fonds pour la Formation \`a la Recherche dans l'Industrie et dans l'Agriculture (FRIA) and Fonds voor Wetenschappelijk Onderzoek contract No. 1228724N (Belgium); the Beijing Municipal Science \& Technology Commission, No. Z191100007219010, the Fundamental Research Funds for the Central Universities, the Ministry of Science and Technology of China under Grant No. 2023YFA1605804, the Natural Science Foundation of China under Grant No. 12535004, and USTC Research Funds of the Double First-Class Initiative No.\ YD2030002017 (China); the Ministry of Education, Youth and Sports (MEYS) of the Czech Republic; the Shota Rustaveli National Science Foundation, grant FR-22-985 (Georgia); the Deutsche Forschungsgemeinschaft (DFG), among others, under Germany's Excellence Strategy -- EXC 2121 ``Quantum Universe" -- 390833306, and under project number 400140256 - GRK2497; the Hellenic Foundation for Research and Innovation (HFRI), Project Number 2288 (Greece); the Hungarian Academy of Sciences, the New National Excellence Program - \'UNKP, the NKFIH research grants K 131991, K 133046, K 138136, K 143460, K 143477, K 146913, K 146914, K 147048, 2020-2.2.1-ED-2021-00181, TKP2021-NKTA-64, and 2025-1.1.5-NEMZ\_KI-2025-00004 (Hungary); the Council of Science and Industrial Research, India; ICSC -- National Research Center for High Performance Computing, Big Data and Quantum Computing, FAIR -- Future Artificial Intelligence Research, and CUP I53D23001070006 (Mission 4 Component 1), funded by the NextGenerationEU program, the Italian Ministry of University and Research (MUR) under Bando PRIN 2022 -- CUP I53C24002390006, PRIN PRIMULA 2022RBYK7T (Italy); the Latvian Council of Science; the Ministry of Education and Science, project no. 2022/WK/14, and the National Science Center, contracts Opus 2021/41/B/ST2/01369, 2021/43/B/ST2/01552, 2023/49/B/ST2/03273, and the NAWA contract BPN/PPO/2021/1/00011 (Poland); the Funda\c{c}\~ao para a Ci\^encia e a Tecnologia (Portugal); the National Priorities Research Program by Qatar National Research Fund; MICIU/AEI/10.13039/501100011033, ERDF/EU, ``European Union NextGenerationEU/PRTR", projects PID2022-142604OB-C21, PID2022-139519OB-C21, PID2023-147706NB-I00, PID2023-148896NB-I00, PID2023-146983NB-I00, PID2023-147115NB-I00, PID2023-148418NB-C41, PID2023-148418NB-C42, PID2023-148418NB-C43, PID2023-148418NB-C44, PID2024-158190NB-C22, RYC2021-033305-I, RYC2024-048719-I, CNS2023-144781, CNS2024-154769 and Plan de Ciencia, Tecnolog{\'i}a e Innovaci{\'o}n de Asturias, Spain; the Chulalongkorn Academic into Its 2nd Century Project Advancement Project, the National Science, Research and Innovation Fund program IND\_FF\_68\_369\_2300\_097, and the Program Management Unit for Human Resources \& Institutional Development, Research and Innovation, grant B39G680009 (Thailand); the Eric \& Wendy Schmidt Fund for Strategic Innovation through the CERN Next Generation Triggers project under grant agreement number SIF-2023-004; the Kavli Foundation; the Nvidia Corporation; the SuperMicro Corporation; the Welch Foundation, contract C-1845; and the Weston Havens Foundation (USA).
\end{acknowledgments}

\bibliography{auto_generated}
\cleardoublepage \appendix\section{The CMS Collaboration \label{app:collab}}\begin{sloppypar}\hyphenpenalty=5000\widowpenalty=500\clubpenalty=5000\cmsinstitute{Yerevan Physics Institute, Yerevan, Armenia}
{\tolerance=6000
A.~Gevorgyan\cmsorcid{0000-0003-2751-9489}, A.~Hayrapetyan, V.~Makarenko\cmsorcid{0000-0002-8406-8605}, A.~Tumasyan\cmsAuthorMark{1}\cmsorcid{0009-0000-0684-6742}
\par}
\cmsinstitute{Marietta Blau Institute for Particle Physics, Vienna, Austria}
{\tolerance=6000
W.~Adam\cmsorcid{0000-0001-9099-4341}, L.~Benato\cmsorcid{0000-0001-5135-7489}, T.~Bergauer\cmsorcid{0000-0002-5786-0293}, M.~Dragicevic\cmsorcid{0000-0003-1967-6783}, P.S.~Hussain\cmsorcid{0000-0002-4825-5278}, M.~Jeitler\cmsAuthorMark{2}\cmsorcid{0000-0002-5141-9560}, N.~Krammer\cmsorcid{0000-0002-0548-0985}, A.~Li\cmsorcid{0000-0002-4547-116X}, D.~Liko\cmsorcid{0000-0002-3380-473X}, M.~Matthewman, J.~Schieck\cmsAuthorMark{2}\cmsorcid{0000-0002-1058-8093}, R.~Sch\"{o}fbeck\cmsAuthorMark{2}\cmsorcid{0000-0002-2332-8784}, M.~Shooshtari\cmsorcid{0009-0004-8882-4887}, M.~Sonawane\cmsorcid{0000-0003-0510-7010}, N.~Van~Den~Bossche\cmsorcid{0000-0003-2973-4991}, W.~Waltenberger\cmsorcid{0000-0002-6215-7228}, C.-E.~Wulz\cmsAuthorMark{2}\cmsorcid{0000-0001-9226-5812}
% ...
\par}
\vskip\cmsinstskip
\dag:~Deceased\\
$^{1}$Also at Yerevan State University, Yerevan, Armenia\\
$^{2}$Also at TU Wien, Vienna, Austria\\
% ...
\end{sloppypar}
%%% END EDITABLE REGION %%%
% skeleton_end
\end{document}